\documentclass{article}

\usepackage{nicematrix}

\usepackage{framed}

\usepackage{lipsum}

\usepackage{tikz, pgf, pgfplots}
\usetikzlibrary
	{
		arrows, automata, chains, datavisualization.formats.functions, fadings, fit, positioning, quantikz, quotes, shapes
	}
\usepackage{tikzpeople}
\usepackage{fontawesome5}
\newcounter{mathseed}
\pgfmathsetseed{\arabic{mathseed}}
\pgfdeclarelayer{background}
\pgfsetlayers{background,main}
\pgfdeclaredecoration{irregular fractal line}{init}
{
	\state{init}[width=\pgfdecoratedinputsegmentremainingdistance]
	{
		\pgfpathlineto{\pgfpoint{random*\pgfdecoratedinputsegmentremainingdistance}{(random*\pgfdecorationsegmentamplitude-0.02)*\pgfdecoratedinputsegmentremainingdistance}}
		\pgfpathlineto{\pgfpoint{\pgfdecoratedinputsegmentremainingdistance}{0pt}}
	}
}
\tikzset
{
	paper/.style =
	{
		draw = MyDarkBlue!10, blur shadow, every shadow/.style = { opacity = 1, MyDarkBlue }, shading = bilinear interpolation, lower left = MyDarkBlue!10, upper left = MyDarkBlue!5, upper right = GreenTeal!75, lower right = MyDarkBlue!5, fill=none
	},
	irregular cloudy border/.style =
	{
		decoration = { irregular fractal line, amplitude = 0.2 }, decorate,
	},
	irregular spiky border/.style =
	{
		decoration = { irregular fractal line, amplitude = -0.2 }, decorate,
	},
	ragged border/.style =
	{
		decoration = {random steps, segment length = 7mm, amplitude = 2mm }, decorate
	}
}
\def\tornpaper#1{%
	\ifthenelse{\isodd{\value{mathseed}}}
	{%
		\tikz
		{
			\node[inner sep = 1em] (A) {#1};		
			\begin{pgfonlayer}{background}			
				\fill[paper]						
				\pgfextra{\pgfmathsetseed{\arabic{mathseed}}\addtocounter{mathseed}{1}}%
				{decorate[irregular cloudy border]{decorate{decorate{decorate{decorate[ragged border]{
										(A.north west) -- (A.north east)
				}}}}}}
				-- (A.south east)
				\pgfextra{\pgfmathsetseed{\arabic{mathseed}}}%
				{decorate[irregular spiky border]{decorate{decorate{decorate{decorate[ragged border]{
										-- (A.south west)
				}}}}}}
				-- (A.north west);
			\end{pgfonlayer}
		}
	}
	{%
		\tikz{
			\node[inner sep=1em] (A) {#1};  
			\begin{pgfonlayer}{background}  
				\fill[paper] 
				\pgfextra{\pgfmathsetseed{\arabic{mathseed}}\addtocounter{mathseed}{1}}%
				{decorate[irregular spiky border]{decorate{decorate{decorate{decorate[ragged border]{
										(A.north east) -- (A.north west)
				}}}}}}
				-- (A.south west)
				\pgfextra{\pgfmathsetseed{\arabic{mathseed}}}%
				{decorate[irregular cloudy border]{decorate{decorate{decorate{decorate[ragged border]{
										-- (A.south east)
				}}}}}}
				-- (A.north east);
		\end{pgfonlayer}}
	}
}

\usepackage{amsthm}
\usepackage{amssymb}
\usepackage[bb = boondox]{mathalfa}
\usepackage{dsfont}
\usepackage{newtxmath}
\usepackage{mathtools}
\usepackage{multicol}
\usepackage{braket}
\usepackage{physics}
\numberwithin{equation}{section}

\usepackage{yquant}

\usepackage[a4paper, left = 2.5cm, right = 2.5 cm, top = 2.5cm, bottom = 3.0 cm]{geometry}

\definecolor{MyLightRed}{RGB}{244, 213, 245}
\definecolor{WordRed}{RGB}{255, 0, 102}
\definecolor{RedDarkLightest}{HTML}{ff0088}
\definecolor{RedDarkLight}{HTML}{ea005f}
\definecolor{RedPurple}{HTML}{aa007f}
\definecolor{Purple}{HTML}{911146}
\definecolor{WordLightGreen}{RGB}{140, 214, 192}
\definecolor{WordGreen}{RGB}{0, 176, 80}
\definecolor{GreenLightest}{HTML}{00ffa0}
\definecolor{GreenLighter1}{HTML}{00b383}
\definecolor{GreenLighter2}{HTML}{00aa7f}
\definecolor{GreenDark}{HTML}{225522}
\definecolor{GreenTeal}{HTML}{008080}
\definecolor{WordIceBlue}{RGB}{223, 227, 229}
\definecolor{MyVeryLightBlue}{RGB}{211, 245, 247}
\definecolor{WordBlueVeryLight}{RGB}{0, 176, 240}
\definecolor{WordBlueLight}{RGB}{0, 112, 192}
\definecolor{WordBlueDark}{RGB}{46, 116, 181}
\definecolor{WordBlueDarker}{RGB}{31, 78, 121}
\definecolor{WordBlueDarker25}{RGB}{54, 96, 146}
\definecolor{WordBlueDarker50}{RGB}{36, 64, 98}
\definecolor{WordBlueDarkest}{RGB}{0, 32, 96}
\definecolor{WordBlue}{RGB}{19, 65, 99}
\definecolor{MyBlue}{RGB}{0, 64, 128}
\definecolor{MyDarkBlue}{RGB}{0, 51, 102}
\definecolor{BlueVeryDark}{HTML}{222255}
\definecolor{MagentaVeryLight}{RGB}{178, 162, 201}
\definecolor{MagentaLighter}{RGB}{161, 106, 221}
\definecolor{MagentaLight}{RGB}{128, 100, 162}
\definecolor{MagentaDark}{RGB}{106, 65, 152}
\definecolor{MagentaVeryDark}{RGB}{97, 75, 128}
\definecolor{WordAquaLighter80}{RGB}{218, 238, 243}
\definecolor{WordAquaLighter60}{RGB}{183, 222, 232}
\definecolor{WordAquaLighter40}{RGB}{146, 205, 220}
\definecolor{WordAquaDarker25}{RGB}{49, 134, 155}
\definecolor{WordAquaDarker50}{RGB}{33, 89, 103}
\definecolor{WordVeryLightTeal}{RGB}{223, 236, 235}
\definecolor{WordLightTeal}{RGB}{160, 199, 197}
\definecolor{WordDarkTealLighter80}{RGB}{207, 223, 234}
\definecolor{WordDarkTeal}{RGB}{72, 123, 119}
\definecolor{WordDarkerTeal}{RGB}{48, 82, 80}
\definecolor{WordTurquoiseLighter80}{RGB}{209, 238, 249}
\definecolor{Brown}{HTML}{666633}

\usepackage{xcolor}
\usepackage{varwidth}
\usepackage[skins, breakable, listings, theorems] {tcolorbox}
\usepackage{graphicx}
\usepackage{hyperref}
\hypersetup
{
	colorlinks,
	linkcolor = {WordRed!75!black},
	citecolor = {WordBlueDarker25},
	urlcolor = {WordAquaDarker50}
}
\usepackage{colortbl}
\usepackage{float}
\usepackage{cellspace}
\usepackage{makecell}
\usepackage{multirow}
\usepackage[linesnumbered, tworuled, vlined]{algorithm2e}
\usepackage{appendix}
\usepackage{enumitem}
\usepackage{xintexpr}
\newcommand{\orcidicon}[1]{\href{https://orcid.org/#1}{\includegraphics[height=\fontcharht\font`\B]{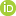}}}

\newtheorem{definition}{Definition}[section]

\newtheorem{example}{Example}[section]

\title
	{
		One-to-Many Simultaneous Secure Quantum Information Transmission
	}

\author
	{
		Theodore Andronikos$^1$\orcidicon{0000-0002-3741-1271}
		and
		Alla Sirokofskich$^2$\\
		\\
		$^1$ \ Department of Informatics, Ionian University, \\
		7 Tsirigoti Square, 49100 Corfu, Greece; \\
		andronikos@ionio.gr \\
		$^2$ \ Department of History and Philosophy of Sciences, \\
		National and Kapodistrian University of Athens, \\
		Athens 15771, Greece; \\
		asirokof@math.uoa.gr
	}

\begin{document}

\maketitle

\begin{abstract}
	This paper presents a new quantum protocol designed to simultaneously transmit information from one source to many recipients. The proposed protocol, which is based on the phenomenon of entanglement, is completely distributed and is provably information-theoretically secure. Numerous existing quantum protocols guarantee secure information communication between two parties but are not amenable to generalization in situations where the source must transmit information to two or more parties, so they must be applied sequentially two or more times in such a setting. The main novelty of the new protocol is its extensibility and generality to situations involving one party that must simultaneously communicate different, in general, messages to an arbitrary number of spatially distributed parties. This is achieved by the special way employed to encode the transmitted information in the entangled state of the system, one of the distinguishing features compared to previous protocols. This protocol can prove expedient whenever an information broker, say, Alice, must communicate distinct secret messages to her agents, all in different geographical locations, in one go. Due to its relative complexity, compared to similar cryptographic protocols, as it involves communication among $n$ parties, and relies on $\ket{ GHZ_{ n } }$ tuples, we provide an extensive and detailed security analysis so as to prove that it is information-theoretically secure. Finally, in terms of its implementation, the prevalent characteristic of the proposed protocol is its uniformity and simplicity because it only requires CNOT and Hadamard gates, and the local quantum circuits are identical for all information recipients.
\end{abstract}
\section{Introduction} \label{sec: Introduction}

In today's world, advocating for the significance of privacy and security in every facet of our lives as individuals hardly needs justification. Privacy is not just a fundamental constitutional right but a cornerstone that demands respect and safeguarding in all circumstances. This imperative has driven the development and deployment of robust technical tools aimed at securing our digital data. The pursuit of foolproof algorithms and protocols to protect our privacy from unauthorized access stands as a prominent theme in current research. However, this endeavor is far from simple, given that we've entered a new scientific epoch, the quantum era, offering the potential of unprecedented computational power. This untapped power introduces novel algorithms that have the potential to compromise the security provided by well-established classical methods. Two illustrative examples underscoring this point are Shor's algorithm \cite{Shor1994} and Grover's algorithm \cite{Grover1996}. Shor's algorithm has the capability to factorize large numbers in polynomial time, posing a practical threat to public key cryptosystems. Grover's algorithm accelerates unordered search tasks and may also be leveraged to attack symmetric key cryptosystems like AES.

As of today, quantum computers with the potential to challenge the classical status quo have not materialized. However, recent remarkable progress, as exemplified by IBM's 127-qubit Eagle processor \cite{IBMEagle} and the more recent 433-qubit Osprey processor \cite{IBMOsprey}, suggests that this may change sooner than initially expected. It appears prudent, if not imperative, to enhance our algorithms and protocols significantly before they become a vulnerability to our security infrastructure. This tremendous effort has given rise to two new scientific fields: post-quantum or quantum-resistant cryptography and quantum cryptography. The former represents an evolutionary step from the current state of affairs \cite{chen2016report, alagic2019status, alagic2020status, alagic2022status}, addressing security concerns by relying on carefully chosen computationally challenging problems, an approach that has proven effective thus far. The latter, quantum cryptography, capitalizes on the laws of nature, such as entanglement, monogamy of entanglement, the no-cloning theorem, and nonlocality, to establish unassailable security.

In our view, the long-term trajectory of cryptography inevitably leads to quantum cryptography, which stands as a pivotal and contemporary research focus. This transition arises from the overwhelming advantages offered by the fundamental properties of quantum mechanics. These properties not only enable the secure protection of information but also facilitate efficient information transmission through the utilization of entangled states, as initially proposed by Arthur Ekert \cite{Ekert1991}. Ekert's groundbreaking E91 quantum key distribution protocol (QKD) demonstrated the feasibility of key distribution using EPR pairs. Following this seminal work by Ekert, the field of quantum cryptography experienced a rapid proliferation of entanglement-based QKD protocols \cite{Bennett1992, Gisin2004, inoue2002differential, guan2015experimental, waks2006security, Ampatzis2021}. This proliferation has underscored the significance of this approach and has spurred the research community to further extend the field by exploring other cryptographic primitives, such as quantum secret sharing. Quantum cryptography harnesses these unique and potent quantum phenomena to design secure protocols for a wide array of critical applications, including key distribution \cite{Bennett1984, Ekert1991, Gisin2004, inoue2002differential, guan2015experimental, waks2006security, Ampatzis2021}, secret sharing \cite{Hillery1999, Ampatzis2022, Ampatzis2023}, quantum teleportation \cite{Bennett1993}, cloud storage \cite{attasena2017secret, ermakova2013secret}, quantum Byzantine Agreement \cite{Andronikos2023a}, and blockchain \cite{cha2021blockchain, Sun2020, Qu2023}.

Another notable research direction in this field is Quantum Secure Direct Communication (QSDC for short) that was initiated in \cite{Long2002}. The most important characteristic of QSDC, which distinguishes it from standard key distribution that establishes a common random key between two parties, is that QSDC transmits information directly and without using an existing key. The classical channel is employed only for detection purposes and not for transmitting information necessary to decipher the secret message. The intended recipient deciphers the secret information after receiving the quantum states via the quantum channel. For a though and comprehensive review of the current state of the field, we refer the reader to the recent \cite{Pan2023}. In a similar vein, the concept of Direct Secure Quantum Communication (DSQC) was initially proposed and further pursued in \cite{Beige2002, Bostroem2002, Nguyen2004} DSQC, also different from quantum key distribution, is designed to transmit a secret message directly without establishing in advance a shared random key to encrypt it. The characteristic trait of DSQC is that in order to decode the secret information, one additional classical bit is required for each qubit. We also mention the important concept of Quantum Private Comparison (QPC), which applies to situations where multiple users who do not trust each other want to conduct secure multi-party computation and obtain the results without revealing their private information. QPC allows all participants to obtain the privacy comparison results at the same time, while ensuring that the privacy information of each participant is confidential and cannot be stolen by other participants. For more details, one may consult the recent \cite{Zhang2023} and references therein.

In this work, we introduce a new entanglement-based protocol for one-to-many simultaneous secure quantum information transmission, or OtMSQIT for short. The characteristic property of the new protocol is its extensibility, as it can be seamlessly generalized to an arbitrary number of entities. The protocol is described as a quantum game, involving the usual suspect Alice. Although, Alice's agents are assumed too many to be named individually, in some small scale examples they are referred to as the usual sidekicks Bob and Charlie. It is expected that the pedagogical nature of games will make the presentation of the technical concepts easier to follow. Quantum games, from their inception in 1999 \cite{Meyer1999,Eisert1999}, have known great acceptance since quantum strategies are sometimes superior to classical ones \cite{Andronikos2018, Andronikos2021, Andronikos2022a}. The famous prisoners' dilemma game provides such the most prominent example, which also applies to other abstract quantum games \cite{Eisert1999,Giannakis2019}, which also applies to other abstract quantum games \cite{Giannakis2015a}. The quantization of many classical systems can even apply to political structures, as was shown in \cite{Andronikos2022}. In the broader context of the use of game-theoretic  While on the subject of games on unconventional environments, it is worth to point out that games in biological systems have recently attracted significant attention \cite{Theocharopoulou2019,Kastampolidou2020a,Kostadimas2021}. It is interesting to observe that biological systems may give rise to biostrategies superior compared to the classical ones, even in the Prisoners' Dilemma iconic game \cite{Kastampolidou2020,Kastampolidou2021,Kastampolidou2023,Papalitsas2021,Adam2023}.

\textbf{Contribution}. This paper presents a new quantum protocol designed to simultaneously transmit information from one source to many recipients. The proposed entanglement-based protocol is completely distributed and is provably information-theoretically secure. Although there many quantum protocols that achieve secure information communication between two parties, most of them are not amenable to generalization to situations where the source must transmit information to two or more recipients in parallel. The main novelty of the new protocol is its extensibility and generality to situations involving one source that must simultaneously communicate different, in general, messages to an arbitrary number of spatially distributed parties. This is achieved by the special way the transmitted information is embedded in the entangled state of the system, one of the distinguishing features compared to previous protocols. This protocol can prove expedient whenever an information broker, say, Alice, must communicate distinct secret messages to a distributed network of agents in one go. Due to its relative complexity, compared to similar cryptographic protocols, as it involves communication among $n$ parties, and relies on $\ket{ GHZ_{ n } }$ tuples, we provide an extensive and detailed security analysis so as to prove that it is information-theoretically secure. In terms of the capabilities of modern quantum apparatus, the implementation of the proposed protocol does not present any difficulty because it only requires CNOT and Hadamard gates. An additional advantage is that the local quantum circuits are identical for all information recipients.

\subsection*{Organization} \label{subsec :Organization}

The paper is organized as follows. Section \ref{sec: Introduction} contains an introduction to the subject along with bibliographic pointers to related works. Section \ref{sec: Background & Terminology} presents the underlying machinery necessary for understanding the technicalities of the protocol. Section \ref{sec: The OtMSSQIT Protocol} provides an analytical and rigorous exposition of the proposed quantum protocol. Section \ref{sec: Security Analysis} is devoted to the detailed security analysis of the protocol, and, finally, Section \ref{sec: Discussion and Conclusions} gives a brief summary of this work, and outlines directions for future research.

\section{Background \& terminology} \label{sec: Background & Terminology}

In the realm of quantum physics, one encounters peculiar hallmark properties that defy classical physics and challenge our everyday intuition. One of the prime examples of this strangeness is entanglement, a phenomenon that not only bewilders but also holds immense potential for accomplishing feats that are difficult or even impossible in the classical world. Entanglement arises in composite quantum systems, typically composed of at least two subsystems, often situated at separate locations. From a mathematical standpoint, a composite system is considered entangled when its state can only be described as a linear combination of two or more product states involving its subsystems. One of the remarkable advantages of quantum entanglement is that when a measurement is performed on one qubit of an entangled pair or tuple, the other qubit(s) instantaneously collapse(s) to the corresponding basis state in the product, regardless of the physical distance separating them. It is precisely this celebrated characteristic of quantum entanglement that finds application in various quantum cryptographic protocols, such as key distribution and secret sharing, among others.

Arguably, the most well-known examples of maximal entanglement are pairs of qubits in one of the four Bell states, also referred to as EPR pairs. For more details, including their precise mathematical description, the interested reader may consult any standard textbook, such as \cite{Nielsen2010, Yanofsky2013a, Wong2022}. Fortunately, maximal entanglement is generalized in the most straightforward and intuitive way in the case of multipartite systems. Perhaps, the most celebrated form of maximal entanglement encountered in composite systems consisting of $n$ qubits, where $n \geq 3$, is the $\ket{ GHZ_{ n } }$ state (GHZ are the initials of the researchers Greenberger, Horne, and Zeilinger). In such a scenario, a composite quantum system consists of $n$ individual qubits, possibly spatially separated, with each qubit considered as a separate subsystem. All these $n$ qubits are entangled in the $\ket{ GHZ_{ n } }$ state, which is mathematically described as follows:

\begin{align} \label{eq: Extended General GHZ_n State}
	\ket{ GHZ_{ n } }
	=
	\frac
	{
		\ket{ 0 }_{ n - 1 } \ket{ 0 }_{ n - 2 } \dots \ket{ 0 }_{ 0 }
		+
		\ket{ 1 }_{ n - 1 } \ket{ 1 }_{ n - 2 } \dots \ket{ 1 }_{ 0 }
	}
	{ \sqrt{ 2 } }
	\ .
\end{align}

In the previous formula \eqref{eq: Extended General GHZ_n State}, the subscript $i, \ 0 \leq i \leq n - 1$, designates the $i^{ th }$ individual qubit. Today, existing quantum computers can produce arbitrary GHZ states using standard quantum gates such as the Hadamard and CNOT gates. Moreover, the circuits that generate these states are very efficient because they require $\lg n$ steps for the $\ket{ GHZ_{ n } }$ state \cite{Cruz2019}.

The protocol introduced in this work requires a more elaborate and general distributed quantum system, in which each individual subsystem is not just a single qubit, but a quantum register $r_{ i }$, $0 \leq i \leq n - 1$, consisting of $m$ qubits. In this respect, the defining property of this setting is that the \emph{corresponding} qubits of all the $n$ registers are entangled in the $\ket{ GHZ_{ n } }$ state. This is formalized by the following Definition \ref{def: Entanglement Distribution Scheme}.

\begin{definition} [Entanglement Distribution Scheme] \label{def: Entanglement Distribution Scheme}
	The $( n, m )$ Symmetric Bit-wise Entanglement Distribution Scheme, denoted by $SBEDS_{ n, m }$, asserts the existence of $n$ spatially distributed quantum registers $r_0, r_{ 1 }, \dots, r_{ n - 1 }$, each containing $m$ bits, satisfying the property that for each $j, 0 \leq j \leq m - 1$, the $n$ qubits occupying the $j^{ th }$ position of each register are entangled in the $\ket{ GHZ_{ n } }$ state.
\end{definition}

As a result, the global state of the composite distributed system is expressed by the next equation, proved in \cite{Ampatzis2023}.

\begin{align} \label{eq: m-Fold Extended General GHZ_n State}
	\ket{ GHZ_{ n } }^{ \otimes m }
	&=
	\frac { 1 } { \sqrt{ 2^{ m } } }
	\sum_{ \mathbf{ x } \in \mathbb{ B }^{ m } }
	\ket{ \mathbf{ x } }_{ n - 1 } \dots \ket{ \mathbf{ x } }_{ 0 }
	\ .
\end{align}

In the above equation \eqref{eq: m-Fold Extended General GHZ_n State}, the following notation is employed.

\begin{itemize}
	\item	$\mathbb{ B }$ stands for $\{ 0, 1 \}$.
	\item	We follow the typical convention of writing bit vectors $\mathbf{ x } \in \mathbb{ B }^{ m }$ in boldface. A bit vector $\mathbf{ x }$ of length $m$ is simply a sequence of $m$ bits $\mathbf{ x } = x_{ m - 1 } \dots x_{ 0 }$. In this fashion, the zero bit vector is designated by $\mathbf{ 0 } = 0 \dots 0$.
	\item	The notation $\mathbf{ x } \in \mathbb{ B }^{ m }$ means that the bit vector $\mathbf{ x }$ ranges through all the $2^{ m }$ bit vector representations of the basis kets.
	\item	To avoid any possible confusion, we use again the indices $i, \ 0 \leq i \leq n - 1$, to make clear that $\ket{ \mathbf{ x } }_{ i }$ denotes the state of the $i^{ th }$ quantum register.
\end{itemize}

A visual depiction of this setup is given in Figure \ref{fig: The Entanglement Distribution Scheme}, where the corresponding qubits comprising the $\ket{ GHZ_{ n } }$ $n$-tuple are drawn with the same color. This composite system contains $m n$ distributed qubits in total because there exist $m$ qubits in each of the $n$ registers. The registers are all assumed to be in different geographic locations, but the entanglement effect due to the $m$ $\ket{ GHZ_{ n } }$ $n$-tuples provides the necessary correlation that enables us to view this as one, albeit distributed, system.

\begin{tcolorbox}
	[
		grow to left by = 0.00 cm,
		grow to right by = 0.00 cm,
		colback = MagentaLight!08,			
		enhanced jigsaw,					
		sharp corners,
		toprule = 1.0 pt,
		bottomrule = 1.0 pt,
		leftrule = 0.1 pt,
		rightrule = 0.1 pt,
		sharp corners,
		center title,
		fonttitle = \bfseries
	]
	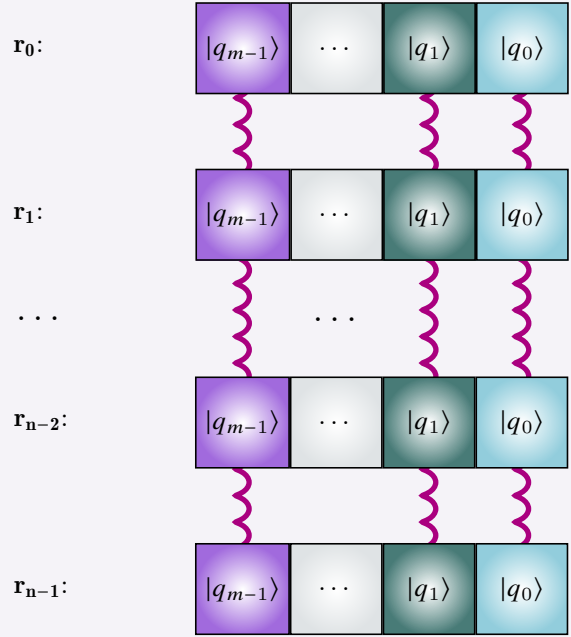
\begin{figure}[H]
		\centering
		\begin{tikzpicture} [ scale = 0.30 ]
			\node
				[
					anchor = center, shade, top color = GreenTeal, bottom color = black, rectangle, text width = 12.00 cm, align = center
				]
			(Label)
			{ \color{white} \textbf{A distributed system consisting of $n$ spatially separated quantum registers $r_{ 0 }, \dots, r_{ n - 1 }$. Each register has $m$ qubits and the corresponding qubits are entangled in the $\ket{ GHZ_{ n } }$ state.} };
			\node [ anchor = west, below left = 1.75 cm and - 3.00 cm of Label ] (QR_n_1) { $\mathbf{ r_{ 0 } }$: \phantom{--} };
			\matrix
				[
					right = 4.00 cm of QR_n_1, anchor = center, column sep = 0.00 mm, row sep = 0.0 mm,
					nodes = { draw = black, minimum size = 12 mm, semithick } 
				]
			{
				\node [ shade, outer color = MagentaLighter, inner color = white ] (QR_n_1-m-1) { $\ket{ q_{ m - 1 } }$ }; &
				\node [ shade, outer color = WordIceBlue, inner color = white ] { \dots }; &
				\node [ shade, outer color = WordDarkTeal, inner color = white ] (QR_n_1-1) { $\ket{ q_{ 1 } }$ }; &
				\node [ shade, outer color = WordAquaLighter40, inner color = white ] (QR_n_1-0) { $\ket{ q_{ 0 } }$ };
				\\
			};
			\node [ anchor = west, below = 1.75 cm of QR_n_1 ] (QR_n_2) { $\mathbf{ r_{ 1 } }$: \phantom{--} };
			\matrix
				[
					right = 4.00 cm of QR_n_2, anchor = center, column sep = 0.00 mm, row sep = 0.0 mm,
					nodes = { draw = black, minimum size = 12 mm, semithick } 
				]
			{
				\node [ shade, outer color = MagentaLighter, inner color = white ] (QR_n_2-m-1) { $\ket{ q_{ m - 1 } }$ }; &
				\node [ shade, outer color = WordIceBlue, inner color = white ] { \dots }; &
				\node [ shade, outer color = WordDarkTeal, inner color = white ] (QR_n_2-1) { $\ket{ q_{ 1 } }$ }; &
				\node [ shade, outer color = WordAquaLighter40, inner color = white ] (QR_n_2-0) { $\ket{ q_{ 0 } }$ };
				\\
			};
			\node [ anchor = west, below = 1.00 cm of QR_n_2 ] (Dots) { \Large \dots };
			\matrix
				[
					right = 3.45 cm of Dots, anchor = center, column sep = 0.00 mm, row sep = 0.0 mm,
					nodes = { minimum size = 14 mm, semithick }
				]
			{
				\node { }; &
				\node { }; &
				\node { \Large \dots }; &
				\node { }; &
				\node { };
				\\
			};
			\node [ anchor = west, below = 1.00 cm of Dots ] (QR_1) { $\mathbf{ r_{ n - 2 } }$: };
			\matrix
				[
				right = 4.00 cm of QR_1, anchor = center, column sep = 0.00 mm, row sep = 0.0 mm,
				nodes = { draw = black, minimum size = 12 mm, semithick }
				]
			{
				\node [ shade, outer color = MagentaLighter, inner color = white ] (QR_1-m-1) { $\ket{ q_{ m - 1 } }$ }; &
				\node [ shade, outer color = WordIceBlue, inner color = white ] { \dots }; &
				\node [ shade, outer color = WordDarkTeal, inner color = white ] (QR_1-1) { $\ket{ q_{ 1 } }$ }; &
				\node [ shade, outer color = WordAquaLighter40, inner color = white ] (QR_1-0) { $\ket{ q_{ 0 } }$ };
				\\
			};
			\node [ anchor = west, below = 1.75 cm of QR_1 ] (QR_0) { $\mathbf{ r_{ n - 1 } }$: };
			\matrix
				[
					right = 4.00 cm of QR_0, anchor = center, column sep = 0.00 mm, row sep = 0.0 mm,
					nodes = { draw = black, minimum size = 12 mm, semithick }
				]
			{
				\node [ shade, outer color = MagentaLighter, inner color = white ] (QR_0-m-1) { $\ket{ q_{ m - 1 } }$ }; &
				\node [ shade, outer color = WordIceBlue, inner color = white ] { \dots }; &
				\node [ shade, outer color = WordDarkTeal, inner color = white ] (QR_0-1) { $\ket{ q_{ 1 } }$ }; &
				\node [ shade, outer color = WordAquaLighter40, inner color = white ] (QR_0-0) { $\ket{ q_{ 0 } }$ };
				\\
			};
			\scoped [ on background layer ]
			\draw
			[ RedPurple, -, >=stealth, line width = 0.7 mm , decoration = coil, decorate ]
			(QR_n_1-m-1.center) -- (QR_n_2-m-1.center) -- (QR_1-m-1.center) -- (QR_0-m-1.center);
			\scoped [ on background layer ]
			\draw
			[ RedPurple, -, >=stealth, line width = 0.7 mm , decoration = coil, decorate ]
			(QR_n_1-1.center) -- (QR_n_2-1.center) -- (QR_1-1.center) -- (QR_0-1.center);
			\scoped [ on background layer ]
			\draw
			[ RedPurple, -, >=stealth, line width = 0.7 mm , decoration = coil, decorate ]
			(QR_n_1-0.center) -- (QR_n_2-0.center) -- (QR_1-0.center) -- (QR_0-0.center);
			\node [ anchor = west, below = 0.75 cm of QR_0 ] (PhantomNode) { };
		\end{tikzpicture}
		\caption{In the above figure, qubits that belong to the same $\ket{ GHZ_{ n } }$ $n$-tuple are drawn with the same color.}
		\label{fig: The Entanglement Distribution Scheme}
	\end{figure}
\end{tcolorbox}

\begin{example} [Alice, Bob \& Charlie] \label{xmp: Alice, Bob & Charlie}
Let us consider a special case of the general setting, featuring the $3$ prolific players Alice, Bob, and Charlie. They are all in different geographical locations, and they possess their own local quantum registers. Moreover, each register contains $9$ qubits. According to the $SBEDS_{ 3, 9 }$ entanglement distribution scheme, there are nine triplets of qubits, and in each triplet the qubits of are entangled in the $\ket{ GHZ_{ 3 } }$ state. The resulting setting is shown in Figure \ref{fig: Alice, Bob, and Charlie's Entangled Quantum Registers}.
\hfill $\triangleleft$
\end{example}

\begin{tcolorbox}
	[
		grow to left by = 0.00 cm,
		grow to right by = 0.00 cm,
		colback = MagentaLight!08,			
		enhanced jigsaw,					
		sharp corners,
		toprule = 1.0 pt,
		bottomrule = 1.0 pt,
		leftrule = 0.1 pt,
		rightrule = 0.1 pt,
		sharp corners,
		center title,
		fonttitle = \bfseries
	]
	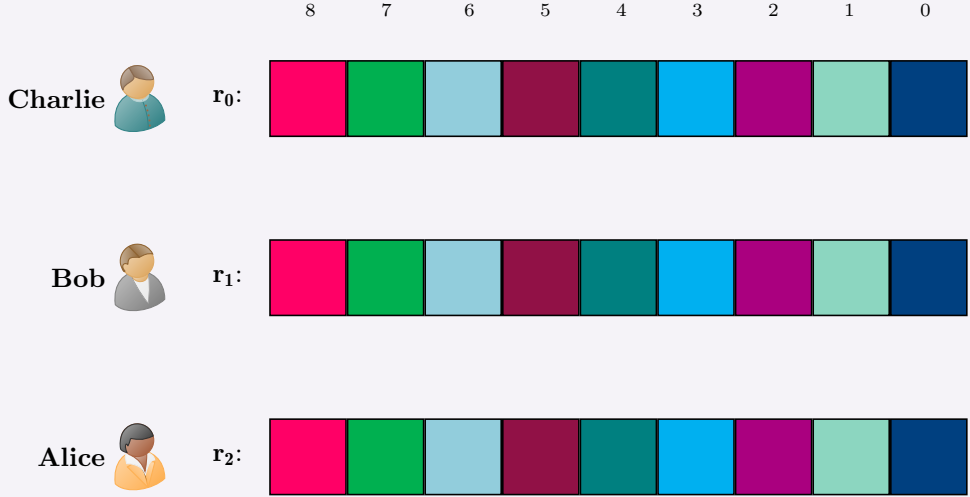
\begin{figure}[H]
		\centering
		\begin{tikzpicture} [scale = 0.75]
			\node
				[
					charlie,
					scale = 1.75,
					anchor = center,
					label = { [ label distance = 0.00 cm ] west: \textbf{Charlie} }
				]
			(Charlie) { };
			\matrix
				[
					column sep = 0.000 mm, above right = 0.10 cm and 1.25 cm of Charlie,
					nodes = { fill = none, minimum size = 10 mm, semithick, align = center, font = \scriptsize }
				]
			{
				\node { $\phantom{-} 8$ }; &
				\node { $\phantom{-} 7$ }; &
				\node { $\phantom{--} 6$ }; &
				\node { $\phantom{--} 5$ }; &
				\node { $\phantom{--} 4$ }; &
				\node { $\phantom{--} 3$ }; &
				\node { $\phantom{--} 2$ }; &
				\node { $\phantom{--} 1$ }; &
				\node { $\phantom{--} 0$ }; \\
			};
			\matrix
				[
					column sep = 0.000 mm, right = 1.25 of Charlie,
					nodes = { draw = black, fill = white, minimum size = 10 mm, semithick, font = \small },
					label = { [ label distance = 0.10 cm ] west: $\mathbf{ r_{ 0 } }$: }
				]
			{
				\node [ fill = WordRed ] { }; &
				\node [ fill = WordGreen ] { }; &
				\node [ fill = WordAquaLighter40 ] { }; &
				\node [ fill = Purple ] { }; &
				\node [ fill = GreenTeal ] { }; &
				\node [ fill = WordBlueVeryLight ] { }; &
				\node [ fill = RedPurple ] { }; &
				\node [ fill = WordLightGreen ] { }; &
				\node [ fill = MyBlue ] { }; \\
			};
			\node
				[
					bob,
					scale = 1.75,
					anchor = center,
					below = 1.50 cm of Charlie,
					label = { [ label distance = 0.00 cm ] west: \textbf{Bob} }
				]
			(Bob) { };
			\matrix
				[
					column sep = 0.000 mm, right = 1.25 of Bob,
					nodes = { draw = black, fill = white, minimum size = 10 mm, semithick, font = \small },
					label = { [ label distance = 0.10 cm ] west: $\mathbf{ r_{ 1 } }$: }
				]
			{
				\node [ fill = WordRed ] { }; &
				\node [ fill = WordGreen ] { }; &
				\node [ fill = WordAquaLighter40 ] { }; &
				\node [ fill = Purple ] { }; &
				\node [ fill = GreenTeal ] { }; &
				\node [ fill = WordBlueVeryLight ] { }; &
				\node [ fill = RedPurple ] { }; &
				\node [ fill = WordLightGreen ] { }; &
				\node [ fill = MyBlue ] { }; \\
			};
			\node
				[
					alice,
					scale = 1.75,
					anchor = center,
					below = 1.50 cm of Bob,
					label = { [ label distance = 0.00 cm ] west: \textbf{Alice} }
				]
			(Alice) { };
			\matrix
				[
					column sep = 0.000 mm, right = 1.25 of Alice,
					nodes = { draw = black, fill = white, minimum size = 10 mm, semithick, font = \small },
					label = { [ label distance = 0.10 cm ] west: $\mathbf{ r_{ 2 } }$: }
				]
			{
				\node [ fill = WordRed ] { }; &
				\node [ fill = WordGreen ] { }; &
				\node [ fill = WordAquaLighter40 ] { }; &
				\node [ fill = Purple ] { }; &
				\node [ fill = GreenTeal ] { }; &
				\node [ fill = WordBlueVeryLight ] { }; &
				\node [ fill = RedPurple ] { }; &
				\node [ fill = WordLightGreen ] { }; &
				\node [ fill = MyBlue ] { }; \\
			};
		\end{tikzpicture}
		\caption{This figure is a pictorial representation of the setting outlined in this example. The $3$ spatially separated players, Alice, Bob, and Charlie possess the $3$ quantum registers $\mathbf{ r_{ 2 } }$, $\mathbf{ r_{ 1 } }$, and $\mathbf{ r_{ 0 } }$, respectively, each containing $9$ qubits. The $3$ qubits occupying the $j^{ th }$ position of each register, $0 \leq j \leq 8$, are entangled in the $\ket{ GHZ_{ 3 } }$ state. To visually indicate this fact, we have drawn the qubits of the same quadruple with the same color.}
		\label{fig: Alice, Bob, and Charlie's Entangled Quantum Registers}
	\end{figure}
\end{tcolorbox}

In addition to $\ket{ GHZ_{ n } }$ tuples, our communication scheme makes use of two other signature states, namely $\ket{ + }$ and $\ket{ - }$, defined as

\begin{tcolorbox}
	[
		grow to left by = 0.00 cm,
		grow to right by = 0.00 cm,
		colback = white,			
		enhanced jigsaw,			
		sharp corners,
		toprule = 0.1 pt,
		bottomrule = 0.1 pt,
		leftrule = 0.1 pt,
		rightrule = 0.1 pt,
		sharp corners,
		center title,
		fonttitle = \bfseries
	]
	\begin{minipage}[b]{0.45 \textwidth}
		\begin{align} \label{eq: Ket +}
			\ket{ + } = H \ket{ 0 } = \frac { \ket{ 0 } + \ket{ 1 } } { \sqrt{ 2 } }
		\end{align}
	\end{minipage} 
	\hfill
	\begin{minipage}[b]{0.45 \textwidth}
		\begin{align} \label{eq: Ket -}
			\ket{ - } = H \ket{ 1 } = \frac { \ket{ 0 } - \ket{ 1 } } { \sqrt{ 2 } }
		\end{align}
	\end{minipage}
\end{tcolorbox}

During the formal mathematical analysis of the proposed protocol, it will be necessary to apply the important and useful formula that expresses the $m$-fold Hadamard transform of an arbitrary basis ket. This formula, proved in most standard textbooks, such as \cite{Nielsen2010} and \cite{Mermin2007}, is given below.

\begin{align} \label{eq: Hadamard m-Fold Ket x}
	H^{ \otimes m } \ket{ \mathbf{ x } }
	&=
	\frac { 1 } { \sqrt{ 2^{ n } } }
	\sum_{ \mathbf{ z } \in \mathbb{ B }^{ m } }
	( - 1 )^{ \mathbf{ z \cdot x } } \ket{ \mathbf{ z } }
	\ .
\end{align}

In \eqref{eq: Hadamard m-Fold Ket x}, the symbolism $\mathbf{ x \cdot y }$ denotes the \emph{inner product modulo} $2$ operation. Given bit vectors $\mathbf{ x }, \mathbf{ y } \in \mathbb{ B }^{ m }$, with $\mathbf{ x } = x_{ m - 1 } \dots x_{ 0 }$ and $\mathbf{ y } = y_{ m - 1 } \dots y_{ 0 }$, $\mathbf{ x \cdot y }$ is defined as

	\begin{align} \label{eq: Inner Product Modulo $2$}
	\mathbf{ x \cdot y }
	&=
	x_{ n - 1 } y_{ n - 1 }
	\oplus \dots \oplus
	x_{ 0 } y_{ 0 }
	\ ,
\end{align}

where $\oplus$ stands for \emph{addition modulo} $2$. The inner product modulo $2$ operation satisfies the following characteristic property. If $\mathbf{ c } \in \mathbb{ B }^{ m }$ is different from $\mathbf{ 0 }$, then for half of the elements $\mathbf{ x } \in \mathbb{ B }^{ m }$, the result of the operation $\mathbf{c} \cdot \mathbf{ x }$ is $0$, and for the remaining half, the result of the operation $\mathbf{ c } \cdot \mathbf{ x }$ is $1$. Obviously, if $\mathbf{ c } = \mathbf{ 0 }$, then for all $\mathbf{ x } \in  \mathbb{ B }^{ m }$, $\mathbf{ c } \cdot \mathbf{ x } = 0$ (a more detailed analysis can be found in \cite{Ampatzis2023}). For future reference, this property is referred to as the characteristic inner product (CIP) property.

\begin{tcolorbox}
	[
		grow to left by = 1.50 cm,
		grow to right by = 1.50 cm,
		colback = white,			
		enhanced jigsaw,			
		sharp corners,
		toprule = 0.1 pt,
		bottomrule = 0.1 pt,
		leftrule = 0.1 pt,
		rightrule = 0.1 pt,
		sharp corners,
		center title,
		fonttitle = \bfseries
	]
	\begin{minipage}[c]{0.485 \textwidth}
	{\small
		\begin{align}
			\mathbf{ c } = \mathbf{ 0 }
			\Rightarrow
			\left\{
			\text{for all } 2^{ m } \text{ bit vectors } \mathbf{ x } \in \mathbb{ B }^{ m },
			\text{ } \mathbf{ c } \cdot \mathbf{ x } = 0
			\right\}
			\nonumber
		\end{align}
	}
	\end{minipage} 
	\hfill
	\begin{minipage}[c]{0.50 \textwidth}
	{\small
		\begin{align} \label{eq: Inner Product Modulo $2$ Property}
			\mathbf{ c } \neq \mathbf{ 0 }
			\Rightarrow
			\left\{
			\begin{matrix*}[l]
				\text{for } 2^{ m - 1 } \text{ bit vectors } \mathbf{ x } \in \mathbb{ B }^{ m }, \ \mathbf{ c } \cdot \mathbf{ x } = 0
				\\
				\text{for } 2^{ m - 1 } \text{ bit vectors } \mathbf{ x } \in \mathbb{ B }^{ m }, \ \mathbf{ c } \cdot \mathbf{ x } = 1
			\end{matrix*}
			\right\}
			\tag{CIP}
		\end{align}
	}
	\end{minipage}
\end{tcolorbox}

As a final note, let us clarify that measurements are performed with respect to the computational basis $\{ \ket{ 0 }, \ket{ 1 } \}$, unless otherwise specified. During the implementation of our protocol, when performing the first validation test, it will also be necessary to make measurements with respect to the Hadamard basis $\{ \ket{ + }, \ket{ - } \}$. Whenever such an occasion arises, it will be mentioned explicitly.

\section{The One-to-Many Simultaneous Secure Quantum Information Transmission Protocol} \label{sec: The OtMSSQIT Protocol}

This section contains an in-depth presentation of the entanglement-based protocol for the one-to-many simultaneous secure quantum information transmission, abbreviated to OtMSQIT from now on. The presentation has the form of a quantum game, involving $n$ players. One of them, is the famous spymaster Alice, who must simultaneously communicate to each of her $n - 1$ agents a secret message. In the general exposition of the game, we refer collectively to the $n - 1$ who remain anonymous. In the examples, where the game is played by a small number of players, namely $3$ or $4$, Alice's agents are the equally prominent heroes Bod, Charlie, and Dave. The secret messages are generally different for every agent, although it is conceivable that in special cases all the messages are identical. The messages themselves may encode secret commands, or encryption keys, or some other type of instruction. Their exact purpose is not important; the crucial thing is that the whole process be information-theoretically secure, so as to ensure that Eve, the adversary who eavesdrops, will not obtain any secret information. The most Eve can do is to obstruct the execution of the OtMSQIT protocol, but even in this case, she will be detected and the protocol will be aborted before the final decryption takes place. The envisioned situation is specified by the next Definition \ref{def: One to Many Simultaneous Secure Quantum Information Transmission}.

\begin{definition} [One to Many Simultaneous Secure Quantum Information Transmission] \label{def: One to Many Simultaneous Secure Quantum Information Transmission}
	Consider the following situation.
	\begin{itemize}
		\item	Alice controls a network of $n - 1$ agents: Agent$_{ 0 }$, \dots, Agent$_{ n - 2 }$. Alice and all her agents reside in different geographical locations.
		\item	Alice must transmit to each of her agents a \emph{personalized information bit vector}, abbreviated to PIV from now on, $\mathbf{ i }_{ k }, \ 0 \leq k \leq n - 2$.
		\item	Time is of the essence, so, to speed things up, Alice wants the information transmission to her agents to take place \emph{simultaneously}, in one go.
		\item	Given the PIVs $\mathbf{ i }_{ 0 }, \dots, \mathbf{ i }_{ n - 2 }$, Alice constructs the \emph{aggregated information bit vector}, AIV from now on, $\mathbf{ i }$ as their concatenation.
		\item	Most importantly, the communication must be \emph{information-theoretically secure}, so that her adversary, the eavesdropper Eve, can't obtain any secret information.
	\end{itemize}
	The task at hand is to come up with a quantum protocol that will provably guarantee that Alice attains all the above goals.
\end{definition}

Let us make some clarifications, to eliminate any possible misunderstanding.

\begin{itemize}
	\item	Theoretically, the number $n$ of players is totally arbitrary, i.e., it may be any large integer. The only conceivable limitation could be the ability of our currently available apparatus to generate $\ket{ GHZ_{ n } }$ tuples when $n$ goes beyond a certain limit.
	\item	Alice assigns a specific ordering to her network of agents. The position $i, \ 0 \leq i \leq n - 2$, of each agent in this ordering is common knowledge, that is Alice and all her agents know who is Agent$_0$, \dots, Agent$_{ n - 2 }$.
	\item	In general, the PIVs are assumed to be of different lengths. This is more realistic and flexible than assuming PIVs of the same length. Obviously, our protocol can easily handle the special case where the information bit vectors have a fixed length.
	\item	Alice communicates via the classical channel to all of her agents the length of the AIV and the lengths $| \mathbf{ i }_{ 0 } |, \dots, | \mathbf{ i }_{ n - 2 } |$ of the PIVs. This does not compromise secrecy because knowing the length of a secret vector does not reveal its contents. We use the symbolism $| \cdot |$ to designate the length, i.e., number of bits, of the enclosed bit vector.
\end{itemize}

We make the important remark that in the construction of the AIV, the order with which PIVs are concatenated is in accordance with the ordering depicted in Figure \ref{fig: The Entanglement Distribution Scheme}. This is because for consistency we adhere to the Qiskit \cite{Qiskit2023} convention in the ordering of qubits, by placing the least significant qubit at the top of the figure and the most significant at the bottom. To rigorously define the AIV, we must first define an auxiliary sequence of positive integers as follows:

\begin{align} \label{eq: Sequence of Lengths}
	m_{ 0 }
	=
	| \mathbf{ i }_{ 0 } |,
	\
	m_{ 1 }
	=
	| \mathbf{ i }_{ 1 } | + | \mathbf{ i }_{ 0 } |,
	\dots,
	m_{ n - 3 }
	=
	| \mathbf{ i }_{ n - 3 } | + \cdots + | \mathbf{ i }_{ 0 } |,
	\
	m
	=
	| \mathbf{ i }_{ n - 2 } | + \cdots + | \mathbf{ i }_{ 0 } |
	\ ,
\end{align}

which allows us to proceed to the following definition of the AIV $\mathbf{ i }$.

\begin{align} \label{eq: The Aggregated Information Bit Vector}
	\mathbf{ i }
	=
	i_{ m - 1 } \cdots i_{ 0 }
	=
	\underbrace { i_{ m - 1 } \cdots i_{ m_{ n - 3 } } }_{ \mathbf{ i_{ n - 2 } } }
	\
	\underbrace { i_{ m_{ n - 3 - 1 } } \cdots i_{ m_{ n - 4 } } }_{ \mathbf{ i }_{ n - 3 } } \
	\dots
	\underbrace { i_{ m_{ 1 } - 1 } \cdots i_{ m_{ 0 } } }_{ \mathbf{ i }_{ 1 } }
	\
	\underbrace { i_{ m_{ 0 } - 1 } \cdots i_{ 0 } }_{ \mathbf{ i }_{ 0 } }
	\ .
\end{align}

From now on, and in accordance with the previous equation \eqref{eq: The Aggregated Information Bit Vector}, we will use $m$ to designate the length of the AIV.

\subsection{Entanglement distribution \& validation stage} \label{subsec: Entanglement Distribution & Validation Stage}

It is helpful to describe the evolution of the OtMSQIT protocol in stages. The first is the entanglement distribution \& validation stage, during which the following tasks take place.

\begin{enumerate} [ left = 1.00 cm, labelsep = 1.50 cm ]
	\renewcommand\labelenumi{(\textbf{EDV}$_\theenumi$)}
	\item	Alice prepares a sequence of $m$ $\ket{ GHZ_{ n } }$ tuples, that is $m n$ qubits, called the \emph{information sequence} IS, which will be used for the actual transmission of the AIV.
	\item	Additionally, Alice prepares the \emph{decoy sequence} DS consisting of $d$ nonentangled $n$-tuples, called \emph{decoy tuples}, which will be used during the first stage of the protocol for the \emph{validation test}. In a decoy tuple, each qubit is prepared in a state that is chosen randomly and with equal probability from the states $\{ \ket{ + }, \ket{ - } \}$. It is important to emphasize that each qubit of the decoy tuple is prepared independently of the other qubits of the same tuple. Altogether, $d n$ decoy qubits will be prepared in the Hadamard basis.
	\item	Assuming that in each $n$-tuple the qubits are numbered from $0$ (the least significant) to $n - 1$ (the most significant), Alice
	\begin{itemize}
		\item[$\diamond$]	stores in her input register, denoted by $AIR$ in Figure \ref{fig: The Quantum Circuit for the OtMSQIT Protocol}, the $( n - 1 )^{ th }$ qubit of each of the
		$m + d$ in total $n$-tuples, and
		\item[$\diamond$]	sends to Agent$_i$ the $i^{ th }$ qubit, $0 \leq i \leq n - 2$, of each of the
		$m + d$ tuples through the quantum channel. These qubits will populate Agent$_i$'s input register, designated by $IR_{ i }$ in Figure \ref{fig: The Quantum Circuit for the OtMSQIT Protocol}. Overall, Alice prepares
		$( m + d ) n$
		qubits, and transmits
		$( m + d ) ( n - 1 )$ qubits to her agents, out of which the
		$m ( n - 1 )$ are information carriers and the
		$d ( n - 1 )$ are decoys.
	\end{itemize}
	\item	It is of critical importance that Alice inserts
	the decoy sequence randomly and uniformly within the information sequence, using an appropriate probability distribution. Obviously, Alice must keep track of the positions of
	decoy tuples. Moreover, for each decoy tuple, Alice must record the states of all of its qubits.
	\item	After the distribution of the
	$m + d$ tuples has been completed, Alice proceeds to conduct the
	validation test, which is analyzed in detail in Section \ref{sec: Security Analysis}. During this test, the $d$ decoy tuples are measured and consumed. If the outcome of the test is deemed a success, Alice knows that her adversary Eve did not manage to tamper with the distribution of the entangled qubits. Thus, the OtMSQIT protocol can safely proceeds to the next stage, in which only the
	$m$ $\ket{ GHZ_{ n } }$ tuples are used. If the outcome of the test considered a failure, the execution of the protocol is aborted.
\end{enumerate}

Let us point out that the case where the protocol is aborted indicates that the security measures are not up to the task at hand. Hence, measures must be taken to enhance security, before the process can start all over again. We also emphasize that in the mathematical analysis of the OtMSQIT protocol and the forthcoming figures, we have intentionally omitted
the decoy tuples in order to streamline and simplify the computation, and to avoid the overcluttering of the figures. Of course, the utilization of these tuples in the
validation test is thoroughly explained in Section \ref{sec: Security Analysis}.

\subsection{Secret embedding stage} \label{subsec: Secret Embedding Stage}

During this stage the AIV is embedded into the entanglement. Alice, using her local quantum circuit, will distribute the information she wants to communicate to her agents into the entangled input registers. At this stage, each input register contains
$m$ qubits, since the $d$ decoy tuples have been previously consumed.
Alice and her $n - 1$ agents, all in different geographical locations, operate on their local quantum circuits. Alice's circuit consists of her input register $AIR$ with $m$ qubits and her output register $AOR$ with just one qubit in the $\ket{ - }$ state, upon which she acts via unitary transforms. All agents have identical local circuits, comprised of the $m$-qubit input registers $IR_{ i }, \ 0 \leq i \leq n - 2$, respectively, on which they apply the $m$-fold Hadamard transform. Although the quantum input registers are spatially separated, they constitute one composite distributed quantum circuit because of the strong correlations among their qubits due to the $SBEDS_{ n, m }$ entanglement distribution scheme of Definition \ref{def: Entanglement Distribution Scheme}. The whole setup is shown in Figure \ref{fig: The Quantum Circuit for the OtMSQIT Protocol}. Recall that all quantum circuits in this paper follow the Qiskit \cite{Qiskit2023} convention in the ordering of qubits, by placing the least significant qubit at the top of the figure and the most significant at the bottom.

\begin{tcolorbox}
	[
		grow to left by = 1.00 cm,
		grow to right by = 1.00 cm,
		colback = white,	
		enhanced jigsaw,						
		sharp corners,
		toprule = 1.0 pt,
		bottomrule = 1.0 pt,
		leftrule = 0.1 pt,
		rightrule = 0.1 pt,
		sharp corners,
		center title,
		fonttitle = \bfseries
	]
	\centering
	\begin{figure}[H]
		\begin{tikzpicture}[ scale = 1.00 ]
			\begin{yquant}
				nobit AUX_0_0;
				[ name = AGENT_0 ] qubits { $IR_0$ } IR_AGENT_0;
				nobit AUX_0_1;
				[ name = Dots_0, register/minimum height = 8 mm ] nobit Dots_0;
				[ name = space_0, register/minimum height = 8 mm ] nobit space_0;
				[ name = Dots_n_2, register/minimum height = 8 mm ] nobit Dots_n_2;
				nobit AUX_n_2_0;
				[ name = AGENT_n_2 ] qubits { $IR_{n - 2}$ } IR_AGENT_n_2;
				nobit AUX_n_2_1;
				[ name = space_n_2, register/minimum height = 8 mm ] nobit space_n_2;
				nobit AUX_A_0;
				[ name = Alice ] qubits { $AIR$ } AIR;
				qubit { $AOR$: \ $\ket{ - }$ } AOR;
				nobit AUX_A_1;
				nobit AUX_A_2;
				[ name = Ph0, WordBlueVeryLight, line width = 0.50 mm, label = Initial State ]
				barrier ( - ) ;
				[ draw = RedPurple!50, fill = RedPurple!50, radius = 0.7 cm ] box { \large \sf{U}$_{ f_{ A } }$} (AIR - AOR);
				[ name = Ph1, WordBlueVeryLight, line width = 0.50 mm, label = Phase 1 ]
				barrier ( - ) ;
				[ draw = GreenLighter2!50, fill = GreenLighter2!50, radius = 0.5 cm ] box {\large \sf{H}$^{ \otimes m }$} IR_AGENT_0;
				[ draw = GreenLighter2!50, fill = GreenLighter2!50, radius = 0.5 cm ] box {\large \sf{H}$^{ \otimes m }$} IR_AGENT_n_2;
				[ draw = GreenLighter2!50, fill = GreenLighter2!50, radius = 0.5 cm ] box {\large \sf{H}$^{ \otimes m }$} AIR;
				[ name = Ph2, WordBlueVeryLight, line width = 0.50 mm, label = Phase 2 ]
				barrier ( - ) ;
				[ draw = white, fill = MagentaDark, radius = 0.5 cm ] measure IR_AGENT_0;
				[ draw = white, fill = MagentaDark, radius = 0.5 cm ] measure IR_AGENT_n_2;
				[ draw = white, fill = MagentaDark, radius = 0.5 cm ] measure AIR;
				[ name = Ph3, WordBlueVeryLight, line width = 0.50 mm, label = Measurement ]
				barrier ( - ) ;
				output { $\ket{ \mathbf{ y }_{ 0 } }$ } IR_AGENT_0;
				output { $\ket{ \mathbf{ y }_{ n - 2 } }$ } IR_AGENT_n_2;
				output { $\ket{ \mathbf{ a } }$ } AIR;
				\node [ below = 5.50 cm ] at (Ph0) { $\ket{ \psi_{ 0 } }$ };
				\node [ below = 5.50 cm ] at (Ph1) { $\ket{ \psi_{ 1 } }$};
				\node [ below = 5.50 cm ] at (Ph2) { $\ket{ \psi_{ 2 } }$};
				\node [ below = 5.50 cm ] at (Ph3) { $\ket{ \psi_{ f }}$};
				\node
					[
						charlie,
						scale = 1.50,
						anchor = east,
						left = 0.85 cm of AGENT_0,
						label = { [ label distance = 0.00 cm ] west: Agent$_0$ }
					]
				(Charlie) { };
				\node
					[
						bob,
						scale = 1.50,
						anchor = center,
						left = 0.50 cm of AGENT_n_2,
						label = { [ label distance = 0.00 cm ] west: Agent$_{n - 2}$ }
					]
				(Bob) { };
				\node
					[
						alice,
						scale = 1.50,
						anchor = center,
						left = 0.75 cm of Alice,
						label = { [ label distance = 0.00 cm ] west: Alice }
					]
				() { };
				\begin{scope} [ on background layer ]
					\node [ below = 0.25 cm of Charlie ] { \Large \vdots };
					\node [ right = 0.25 cm of space_0, rectangle, fill = WordTurquoiseLighter80!75, text width = 10.50 cm, align = center, minimum height = 10 mm ] { \bf Spatially Separated };
					\node [ above = 0.50 cm of Bob ] { \Large \vdots };
					\node [ right = 0.25 cm of space_n_2, rectangle, fill = WordTurquoiseLighter80!75, text width = 10.50 cm, align = center, minimum height = 10 mm ] { \bf Spatially Separated };
				\end{scope}
			\end{yquant}
			\scoped [ on background layer ]
			\draw
			[ RedPurple, -, >=stealth, line width = 0.4 mm , decoration = coil, decorate ]
			(Alice.north east) -- (AGENT_n_2.south east);
			\scoped [ on background layer ]
			\draw
			[ RedPurple, -, >=stealth, line width = 0.4 mm , decoration = coil, decorate ]
			(AGENT_n_2.north east) -- (AGENT_0.south east);
		\end{tikzpicture}
		\caption{The above figure shows the quantum circuits employed by Alice and her agents. Although these circuits are spatially separated, they are correlated due to entanglement and constitute a composite system. The state vectors $\ket{ \psi_{ 0 } }$, $\ket{ \psi_{ 1 } }$, $\ket{ \psi_{ 2 } }$, and $\ket{ \psi_{ f } }$ describe the evolution of the distributed system.}
		\label{fig: The Quantum Circuit for the OtMSQIT Protocol}
	\end{figure}
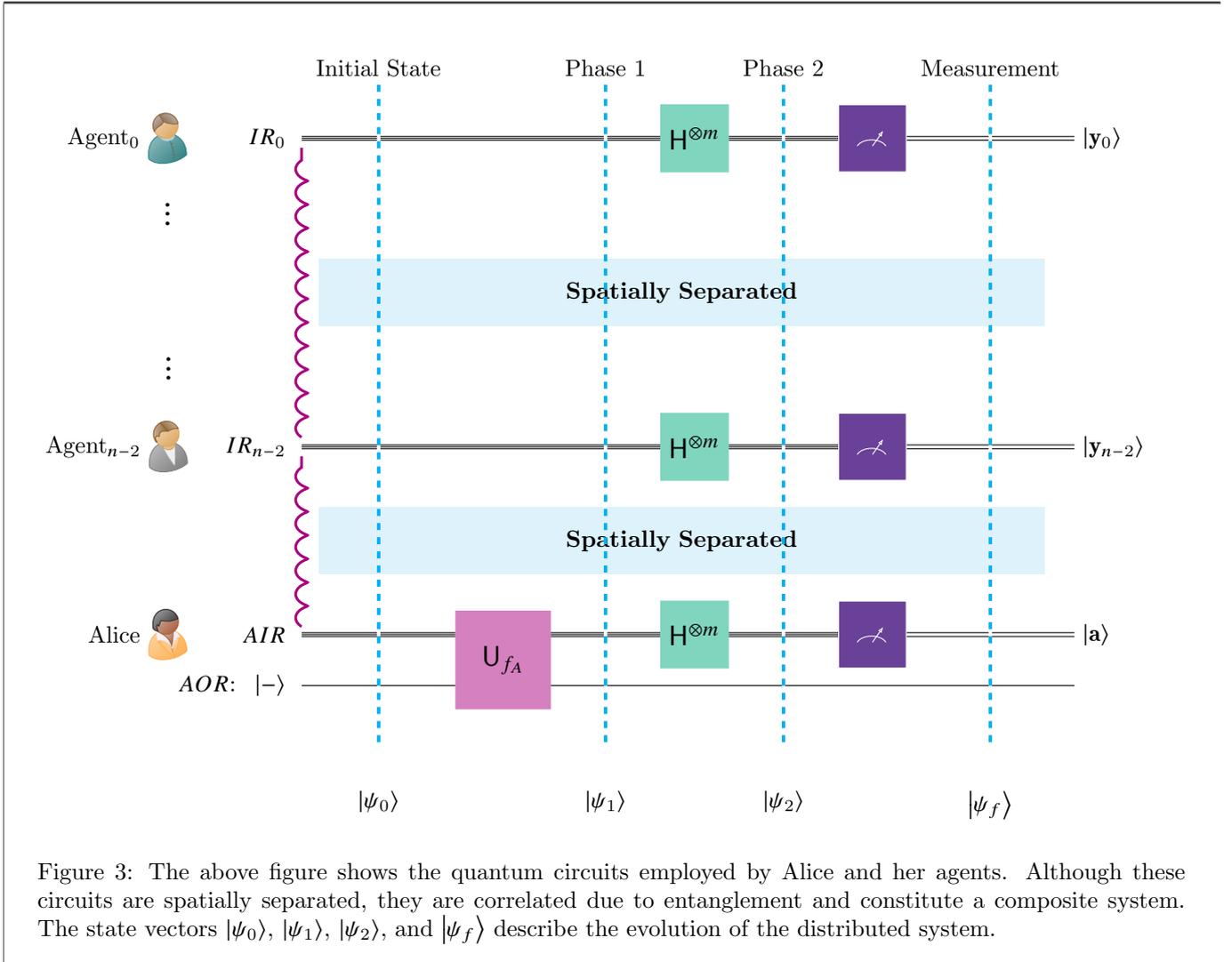
\end{tcolorbox}

The initial state of the distributed quantum circuit (consult Figure \ref{fig: The Quantum Circuit for the OtMSQIT Protocol}) is denoted by $\ket{ \psi_{ 0 } }$. With the help of \eqref{eq: m-Fold Extended General GHZ_n State}, $\ket{ \psi_{ 0 } }$ can be written as

\begin{align} \label{eq: OtMSQIT Protocol Initial State}
	\ket{ \psi_{ 0 } }
	=
	\frac { 1 } { \sqrt{ 2^{ m } } }
	\sum_{ \mathbf{ x } \in \mathbb{ B }^{ m } }
	\ket{ - }_{ A }
	\ket{ \mathbf{ x } }_{ A }
	\ket{ \mathbf{ x } }_{ n - 2 }
	\dots
	\ket{ \mathbf{ x } }_{ 0 }
	\ .
\end{align}

Alice initiates the execution of the OtMSQIT protocol by acting on her local input register $AIR$ via the unitary transform $U_{ f_{ A } }$. By doing so, she embeds the secret information she intends to communicate to her $n - 1$ agents to the distributed circuit. The unitary transform $U_{ f_{ A } }$ is based on the function $f_{ A }$, which uses the AIV $\mathbf{ i }$, as shown below

\begin{align} \label{eq: Alice's Function f_A}
	f_{ A } ( \mathbf{ x } )
	&=
	\mathbf{ i } \cdot \mathbf{ x }
	\ .
\end{align}

The unitary transform $U_{ f_{ A } }$ itself implements the ubiquitous scheme

\begin{align} \label{eq: Alice's Unitary Transform U_f_A}
	U_{ f_{ A } }
	\colon
	\ket{ y }_{ A }
	\ket{ \mathbf{ x } }_{ A }
	\rightarrow
	\ket{ y \oplus f_{ A } ( \mathbf{ x } ) }_{ A }
	\ket{ \mathbf{ x } }_{ A }
	\ .
\end{align}

By combining \eqref{eq: Alice's Function f_A} and \eqref{eq: Alice's Unitary Transform U_f_A}, $U_{ f_{ A } }$ can be explicitly written as

\begin{align} \label{eq: Explicit Alice's Unitary Transform U_f_A}
	U_{ f_{ A } }
	\colon
	\ket{ - }_{ A }
	\ket{ \mathbf{ x } }_{ A }
	\rightarrow
	( - 1 )^{ \mathbf{ i } \cdot \mathbf{ x } }
	\
	\ket{ - }_{ A }
	\ket{ \mathbf{x} }_{ A }
	\ .
\end{align}

The action of the $U_{ f_{ A } }$ drives the system at the end of Phase 1 to state $\ket{ \psi_{ 1 } }$:

\begin{align} \label{eq: OtMSQIT Protocol Phase 1}
	\ket{ \psi_{ 1 } }
	&=
	\frac { 1 } { \sqrt{ 2^{ m } } }
	\sum_{ \mathbf{ x } \in \mathbb{ B }^{ m } }
	\left(
	U_{ f_{ A } }
	\ket{ - }_{ A }
	\ket{ \mathbf{ x } }_{ A }
	\right)
	\
	\ket{ \mathbf{ x } }_{ n - 2 }
	\dots
	\ket{ \mathbf{ x } }_{ 0 }
	\nonumber \\
	&\overset { \eqref{eq: Explicit Alice's Unitary Transform U_f_A} } { = }
	\frac { 1 } { \sqrt{ 2^{ m } } }
	\sum_{ \mathbf{ x } \in \mathbb{ B }^{ m } }
	( - 1 )^{ \mathbf{ i } \cdot \mathbf{ x } }
	\
	\ket{ - }_{ A }
	\ket{ \mathbf{x} }_{ A }
	\ket{ \mathbf{ x } }_{ n - 2 }
	\dots
	\ket{ \mathbf{ x } }_{ 0 }
	\ .
\end{align}

Therefore, at the end of Phase 1, the AIV is embedded in a distributed and implicitly way in the state $\ket{ \psi_{ 1 } }$ of the distributed quantum circuit. The next subsection describes the process by which it can be deciphered by the players.


\subsection{Decryption stage} \label{subsec: Decryption Stage}

The key ingredient in the decryption of the secret is the $m$-fold Hadamard transform that all players apply to their input registers during Phase 2, as visualized in Figure \ref{fig: The Quantum Circuit for the OtMSQIT Protocol}. Hence, at the end of Phase 2 the state of he system has become $\ket{ \psi_{ 2 } }$:

\begin{align} \label{eq: OtMSQIT Protocol Phase 2 - 1}
	\ket{ \psi_{ 2 } }
	=
	\frac { 1 } { \sqrt{ 2^{ m } } }
	\sum_{ \mathbf{ x } \in \mathbb{ B }^{ m } }
	( - 1 )^{ \mathbf{ i } \cdot \mathbf{ x } }
	\
	\ket{ - }_{ A }
	\
	H^{ \otimes m } \ket{ \mathbf{ x } }_{ A }
	\
	H^{ \otimes m } \ket{ \mathbf{ x } }_{ n - 2 }
	\dots
	H^{ \otimes m } \ket{ \mathbf{ x } }_{ 0 }
\end{align}

Using formula \eqref{eq: Hadamard m-Fold Ket x}, $H^{ \otimes m } \ket{ \mathbf{ x } }_{ A }$, $H^{ \otimes m } \ket{ \mathbf{ x } }_{ n - 2 }$, \dots, $H^{ \otimes m } \ket{ \mathbf{ x } }_{ 0 }$ can be rewritten as shown below.

\begin{align*} 
	H^{ \otimes m } \ket{ \mathbf{ x } }_{ A }
	&=
	\frac { 1 } { \sqrt{ 2^{ m } } }
	\sum_{ \mathbf{ a } \in \mathbb{ B }^{ m } }
	( - 1 )^{ \mathbf{ a } \cdot \mathbf{ x } }
	\ket{ \mathbf{ a } }_{ A }
	\\
	H^{ \otimes m } \ket{ \mathbf{ x } }_{ n - 2 }
	&=
	\frac { 1 } { \sqrt{ 2^{ m } } }
	\sum_{ \mathbf{ y }_{ n - 2 } \in \mathbb{ B }^{ m } }
	( - 1 )^{ \mathbf{ y }_{ n - 2 } \cdot \mathbf{ x } }
	\ket{ \mathbf{ y }_{ n - 2 } }_{ n - 2 }
	\\
	&\dots
	\\
	H^{ \otimes m } \ket{ \mathbf{ x } }_{ 0 }
	&=
	\frac { 1 } { \sqrt{ 2^{ m } } }
	\sum_{ \mathbf{ y }_{ 0 } \in \mathbb{ B }^{ m } }
	( - 1 )^{ \mathbf{ y }_{ 0 } \cdot \mathbf{ x } }
	\ket{ \mathbf{ y }_{ 0 } }_{ 0 }
\end{align*}

This allows us to express as $\ket{ \psi_{ 2 } }$ as

{\small
	\begin{align} \label{eq: OtMSQIT Protocol Phase 2 - 2}
		\hspace{- 1.00 cm}
		\ket{ \psi_{ 2 } }
		=
		\frac { 1 } { ( \sqrt{ 2^{ m } )^{ n + 1 } } }
		\sum_{ \mathbf{ a } \in \mathbb{ B }^{ m } }
		\sum_{ \mathbf{ y }_{ n - 2 } \in \mathbb{ B }^{ m } }
		\dots
		\sum_{ \mathbf{ y }_{ 0 } \in \mathbb{ B }^{ m } }
		\sum_{ \mathbf{ x } \in \mathbb{ B }^{ m } }
		( - 1 )^{ ( \mathbf{ i } \oplus \mathbf{ a } \oplus \mathbf{ y }_{ n - 2 } \oplus \dots \oplus \mathbf{ y }_{ 0 } ) \cdot \mathbf{ x } }
		\ket{ - }_{ A }
		\ket{ \mathbf{ a } }_{ A }
		\ket{ \mathbf{ y }_{ n - 2 } }_{ n - 2 }
		\dots
		\ket{ \mathbf{ y }_{ 0 } }_{ 0 }
		\ .
	\end{align}
}

At this point it is expedient to recall the characteristic inner product property \eqref{eq: Inner Product Modulo $2$ Property}. This property implies that whenever $\mathbf{ i } \oplus \mathbf{ a } \oplus \mathbf{ y }_{ n - 2 } \oplus \dots \oplus \mathbf{ y }_{ 0 } \neq \mathbf{ 0 }$, or, equivalently, $\mathbf{ a } \oplus \mathbf{ y }_{ n - 2 } \oplus \dots \oplus \mathbf{ y }_{ 0 } \neq \mathbf{ i }$, the sum $\sum_{ \mathbf{ x } \in \mathbb{ B }^{ m } }$ $( - 1 )^{ ( \mathbf{ i } \oplus \mathbf{ a } \oplus \mathbf{ y }_{ n - 2 } \oplus \dots \oplus \mathbf{ y }_{ 0 } ) \cdot \mathbf{ x } }$ $\ket{ - }_{ A }$ $\ket{ \mathbf{ a } }_{ A }$ $\ket{ \mathbf{ y }_{ n - 2 } }_{ n - 2 }$ $\dots$ $\ket{ \mathbf{ y }_{ 0 } }_{ 0 }$ in \eqref{eq: OtMSQIT Protocol Phase 2 - 2} is just $0$. In contrast, if $\mathbf{ i } \oplus \mathbf{ a } \oplus \mathbf{ y }_{ n - 2 } \oplus \dots \oplus \mathbf{ y }_{ 0 } = \mathbf{ 0 }$, or, equivalently, $\mathbf{ a } \oplus \mathbf{ y }_{ n - 2 } \oplus \dots \oplus \mathbf{ y }_{ 0 } = \mathbf{ i }$, the sum $\sum_{ \mathbf{ x } \in \mathbb{ B }^{ m } }$ $( - 1 )^{ ( \mathbf{ i } \oplus \mathbf{ a } \oplus \mathbf{ y }_{ n - 2 } \oplus \dots \oplus \mathbf{ y }_{ 0 } ) \cdot \mathbf{ x } }$ $\ket{ - }_{ A }$ $\ket{ \mathbf{ a } }_{ A }$ $\ket{ \mathbf{ y }_{ n - 2 } }_{ n - 2 }$ $\dots$ $\ket{ \mathbf{ y }_{ 0 } }_{ 0 }$ is equal to $2^{ m }$ $\ket{ - }_{ A }$ $\ket{ \mathbf{ a } }_{ A }$ $\ket{ \mathbf{ y }_{ n - 2 } }_{ n - 2 }$ $\dots$ $\ket{ \mathbf{ y }_{ 0 } }_{ 0 }$. Thus, $\ket{ \psi_{ 2 } }$ can be cast in the following reduced form:

\begin{align} \label{eq: OtMSQIT Protocol Phase 2 - 3}
	\ket{ \psi_{ 2 } }
	=
	\frac { 1 } { ( \sqrt{ 2^{ m } )^{ n - 1 } } }
	\sum_{ \mathbf{ a } \in \mathbb{ B }^{ m } }
	\sum_{ \mathbf{ y }_{ n - 2 } \in \mathbb{ B }^{ m } }
	\dots
	\sum_{ \mathbf{ y }_{ 0 } \in \mathbb{ B }^{ m } }
	\ket{ - }_{ A }
	\ket{ \mathbf{ a } }_{A}
	\ket{ \mathbf{ y }_{ n - 2 } }_{ n - 2 }
	\dots
	\ket{ \mathbf{ y }_{ 0 } }_{0}
	\ ,
\end{align}

where

\begin{align} \label{eq: Fundamental Correlation Property}
	\mathbf{ a }
	\oplus
	\mathbf{ y }_{ n - 2 }
	\oplus
	\dots
	\oplus
	\mathbf{ y }_{ 0 }
	=
	\mathbf{ i }
	\ .
\end{align}

Following \cite{Ampatzis2023} and \cite{Andronikos2023}, we call equation \eqref{eq: Fundamental Correlation Property} the \textbf{Fundamental Correlation Property} that intertwines Alice and her agents' input registers. This equation has arisen due to the initial entanglement among all the input registers. At the end of Phase 2, the AIV has been embedded in the global state of the distributed quantum circuit and has manifest itself by imposing this constraint upon the contents of the input registers.

Subsequently, Alice and her agents complete the quantum part of the OtMSQIT Protocol by measuring the contents of their input registers in the computational basis, and driving the system to its final state $\ket{ \psi_{ f } }$.

\begin{align}
	\hspace{- 1.00 cm}
	\label{eq: OtMSQIT Protocol Final Measurement}
	\ket{ \psi_{ f } }
	=
	\ket{ - }_{ A }
	\ket{ \mathbf{ a } }_{A}
	\ket{ \mathbf{ y }_{ n - 2 } }_{ n - 2 }
	\dots
	\ket{ \mathbf{ y }_{ 0 } }_{0}
	\ ,
	\text{ where }
	\mathbf{ a },
	\mathbf{ y }_{ n - 2 },
	\dots,
	\mathbf{ y }_{ 0 }
	\in
	\mathbb{ B }^{ m }
	\text{ and }
	\mathbf{ a }
	\oplus
	\mathbf{ y }_{ n - 2 }
	\oplus
	\dots
	\oplus
	\mathbf{ y }_{ 0 }
	=
	\mathbf{ i }
	\ .
\end{align}


We write the contents of Alice and her agents' input registers explicitly as

\begin{align}
	\mathbf{ a }
	&=
	a_{ m - 1 } \cdots a_{ 0 }
	\ , \text{ and }
	\label{eq: Explicit Form of Alice's Input Register}
	\\
	\mathbf{ y }_{ i }
	&=
	y_{ m - 1 }^{ i } \cdots y_{ 0 }^{ i }
	\ , \ 0 \leq i \leq n - 2 \ .
	\label{eq: Explicit Form of Agent$_{ i }$'s Input Register}
\end{align}

Accordingly, we may conceptually divide the AIV and each input register into $n - 1$ segments, so that corresponding segments are correlated to a PIV. We employ the notation $\mathbf{ i }^{ j }$, $\mathbf{ a }^{ j }$, and $\mathbf{ y }_{ i }^{ j }, \ 0 \leq i, j \leq n - 2$, to designate the $j^{ th }$ segment of the AIV, of Alice's input register, and of Agent$_{ i }$'s input register, respectively. The formal definition of segments, which is presented below, relies on the sequence of positive numbers $m_{ 0 }, \dots, m_{ n - 3 }$ that was given in \eqref{eq: Sequence of Lengths}.

\begin{align}
	\mathbf{ i }^{ 0 }
	=
	i_{ m_{ 0 } - 1 } \cdots i_{ 0 }
	\ &, \quad
	\mathbf{ i }^{ j }
	=
	i_{ m_{ j } - 1 } \cdots i_{ m_{ j - 1 } }
	\ , \ 1 \leq j \leq n - 2 \ ,
	\label{eq: Explicit Form of AIV's Segments}
	\\
	\mathbf{ a }^{ 0 }
	=
	a_{ m_{ 0 } - 1 } \cdots a_{ 0 }
	\ &, \quad
	\mathbf{ a }^{ j }
	=
	a_{ m_{ j } - 1 } \cdots a_{ m_{ j - 1 } }
	\ , \ 1 \leq j \leq n - 2 \ , \text{ and }
	\label{eq: Explicit Form of Alice's Segments}
	\\
	\mathbf{ y }_{ i }^{ 0 }
	=
	y_{ m_{ 0 } - 1 }^{ i } \cdots y_{ 0 }^{ i }
	\ &, \quad
	\mathbf{ y }_{ i }^{ j }
	=
	y_{ m_{ j } - 1 }^{ i } \cdots y_{ m_{ j - 1 } }^{ i }
	\ , \ 0 \leq i \leq n - 2 \ , \ 1 \leq j \leq n - 2 \ .
	\label{eq: Explicit Form of Agent$_{ i }$'s Segments}
\end{align}

In view of \eqref{eq: Explicit Form of AIV's Segments} -- \eqref{eq: Explicit Form of Agent$_{ i }$'s Segments}, we may rewrite \eqref{eq: The Aggregated Information Bit Vector}, \eqref{eq: Explicit Form of Alice's Input Register} and \eqref{eq: Explicit Form of Agent$_{ i }$'s Input Register} as

\begin{align}
	\mathbf{ i }
	&=
	\underbrace { \mathbf{ i }^{ n - 2 } }_{ \text{ segment } n - 2 }
	\underbrace { \mathbf{ i }^{ n - 3 } }_{ \text{ segment } n - 3 }
	\dots
	\underbrace { \mathbf{ i }^{ 1 } }_{ \text{ segment } 1 }
	\underbrace { \mathbf{ i }^{ 0 } }_{ \text{ segment } 0 }
	\ ,
	\label{eq: Segment Form of AIV}
	\\
	\mathbf{ a }
	&=
	\underbrace { \mathbf{ a }^{ n - 2 } }_{ \text{ segment } n - 2 }
	\underbrace { \mathbf{ a }^{ n - 3 } }_{ \text{ segment } n - 3 }
	\dots
	\underbrace { \mathbf{ a }^{ 1 } }_{ \text{ segment } 1 }
	\underbrace { \mathbf{ a }^{ 0 } }_{ \text{ segment } 0 }
	\ , \text{ and }
	\label{eq: Segment Form of Alice's Input Register}
	\\
	\mathbf{ y }_{ i }
	&=
	\underbrace { \mathbf{ y }_{ i }^{ n - 2 } }_{ \text{ segment } n - 2 }
	\underbrace { \mathbf{ y }_{ i }^{ n - 3 } }_{ \text{ segment } n - 3 }
	\dots
	\underbrace { \mathbf{ y }_{ i }^{ 1 } }_{ \text{ segment } 1 }
	\underbrace { \mathbf{ y }_{ i }^{ 0 } }_{ \text{ segment } 0 }
	\ , \ 0 \leq i \leq n - 2 \ .
	\label{eq: Segment Form of Agent$_{ i }$'s Input Register}
\end{align}

By combining \eqref{eq: The Aggregated Information Bit Vector}, \eqref{eq: OtMSQIT Protocol Final Measurement}, and \eqref{eq: Segment Form of AIV} -- \eqref{eq: Segment Form of Agent$_{ i }$'s Input Register}, we conclude that

\begin{align} \label{eq: Segment Form of the Fundamental Correlation Property}
	\mathbf{ a }^{ j }
	\oplus
	\mathbf{ y }_{ n - 2 }^{ j }
	\oplus
	\dots
	\oplus
	\mathbf{ y }_{ 0 }^{ j }
	=
	\mathbf{ i }^{ j }
	=
	\mathbf{ i }_{ j }
	\ , \ 0 \leq j \leq n - 2 \ .
	\tag{SCP}
\end{align}

Equation \eqref{eq: Segment Form of the Fundamental Correlation Property} expresses the Fundamental Correlation Property among the $n - 1$ segments, aptly named \textbf{Segment Correlation Property}. This property asserts that by simply XOR-ing the $j^{ th }$ segments of all the input registers, we can recover the PIV $\mathbf{ i }_{ j }$.

From this point onward, the execution of the OtMSQIT protocol will utilize only the classical channel.
For the actual decryption the following transmissions take place through the classical channel.

\begin{enumerate} [ left = 1.00 cm, labelsep = 1.50 cm ]
	\renewcommand\labelenumi{(\textbf{EV}$_\theenumi$)}
	\item	Alice sends to every Agent$_{ i }$, $0 \leq i \leq n - 2$, the $i^{ th }$ segment $\mathbf{ a }^{ i }$ of her input register.
	\item	Agent$_{ i }$, $0 \leq i \leq n - 2$, sends to every other Agent$_{ j }$, $0 \leq j \neq i \leq n - 2$, the $j^{ th }$ segment $\mathbf{ y }_{ i }^{ j }$ of her input register.
\end{enumerate}

Let us emphasize that during the decryption stage

\begin{itemize}
	\item	No agent sends any information to Alice.
	\item	Agent$_{ i }$ keeps to herself the $i^{ th }$ segment $\mathbf{ y }_{ i }^{ i }$ of her input register. Ergo, Eve, despite her knowing the segments $\mathbf{ a }^{ i }$ and $\mathbf{ y }_{ j }^{ i }$, $0 \leq j \neq i \leq n - 2$, transmitted via the classical channel, lacks the crucial ingredient $\mathbf{ y }_{ i }^{ i }$ and is, thus, unable to obtain the PIV $\mathbf{ i }_{ i }$.
\end{itemize}

\begin{example} [Alice, Bob \& Charlie use the OtMSQIT protocol] \label{xmp: Alice, Bob & Charlie Use the OtMSQIT protocol}
	This example features our $3$ protagonists Alice, Bob, and Charlie. As, always they are in different geographical locations, and they possess their own local quantum input registers, each having $6$ qubits. In particular, there are six triplets of qubits, each triplet entangled in the $\ket{ GHZ_{ 3 } }$, according to the $SBEDS_{ 3, 9 }$ entanglement distribution scheme. Alice intends to send the PIVs $\mathbf{ i }_{ B } = 101$ and $\mathbf{ i }_{ C } = 010$ to Bob and Charlie, respectively. This implies that the resulting AIV is $\mathbf{ i } = 101010$, which can be embedded into the global state of the circuit via CNOT gates. The concrete implementation in Qiskit of the general quantum circuit of Figure \ref{fig: The Quantum Circuit for the OtMSQIT Protocol} for this scenario, is visualized in Figure \ref{fig: OtMSQIT Example Quantum Circuit}.

	The final measurements by Alice, Bob and Charlie will produce one of the $2^{ 18 } = 262144$ equiprobable outcomes. Clearly, showing all these outcomes would result in an unintelligible figure, so we have depicted only 25 of them in Figure \ref{fig: OtMSQIT Example Measurement Outcomes}. One may trivially confirm that every outcome satisfies the Segment Correlation Property and verifies equations \eqref{eq: Fundamental Correlation Property} and \eqref{eq: Segment Form of the Fundamental Correlation Property}. Therefore, if Alice and Charlie send their segment $1$ to Bob, then Bob, by XOR-ing with his own segment $1$, will uncover $\mathbf{ i }_{ B } = 101$. Symmetrically, if Alice and Bob send their segment $0$ to Charlie, then Charlie will decipher $\mathbf{ i }_{ C } = 010$.

	To see how this works in practice, let us consider the last bar of the histogram of Figure \ref{fig: OtMSQIT Example Measurement Outcomes}. The label of this bar is $111111 \ 100111 \ 110010$, which, according to the quantum circuit of Figure \ref{fig: The Quantum Circuit for the OtMSQIT Protocol}, means that Alice's input register contains the bit vector $\mathbf{ a } = 111111$, Bob's input register contains the bit vector $\mathbf{ b } = 100111$, and Charlie's input register contains the bit vector $\mathbf{ c } = 110010$. Consequently, Alice, Bob, and Charlie's segments $0$ are $\mathbf{ a }^{ 0 } = 111$, $\mathbf{ b }^{ 0 } = 111$, and $\mathbf{ c }^{ 0 } = 010$, respectively. Alice and Bob communicate their segments $0$ to Charlie, who XORs them with his own segment $0$, i.e., $\mathbf{ a }^{ 0 } \oplus \mathbf{ b }^{ 0 } \oplus \mathbf{ c }^{ 0 } = 111 \oplus 111 \oplus 010 = 010$. By doing so, Charlie retrieves Alice's intended PIV $\mathbf{ i }_{ C } = 010$. Analogously, Alice, Bob, and Charlie's segments $1$ are $\mathbf{ a }^{ 1 } = 111$, $\mathbf{ b }^{ 1 } = 100$, and $\mathbf{ c }^{ 1 } = 110$, respectively. Alice and Charlie communicate their segments $1$ to Bob, who XORs them with his own segment $1$, i.e., $\mathbf{ a }^{ 1 } \oplus \mathbf{ b }^{ 1 } \oplus \mathbf{ c }^{ 1 } = 111 \oplus 100 \oplus 110 = 101$. By doing so, Bob also uncovers Alice's intended PIV $\mathbf{ i }_{ B } = 101$.
	\hfill $\triangleleft$
\end{example}

\begin{tcolorbox}
	[
		grow to left by = 0.00 cm,
		grow to right by = 0.00 cm,
		colback = white,			
		enhanced jigsaw,			
		sharp corners,
		boxrule = 0.01 pt,
		toprule = 0.01 pt,
		bottomrule = 0.01 pt
	]
	\begin{figure}[H]
		\centering
		\includegraphics[ scale = 0.44, angle = 90, trim = {0 0 0cm 0}, clip ]{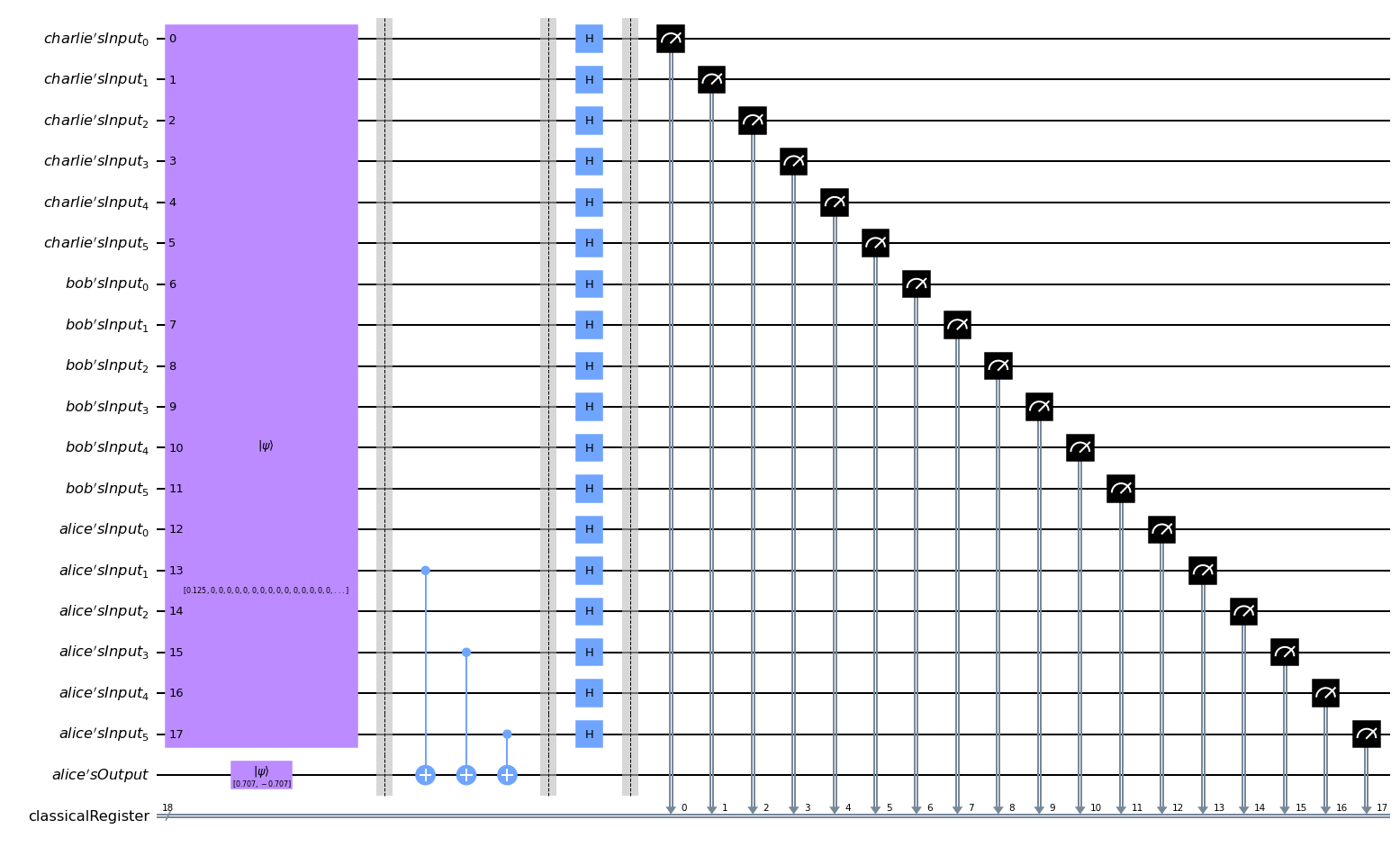}
		\caption{A small scale quantum circuit simulating the OtMSQIT protocol involving Alice and her two agents Bob and Charlie.}
		\label{fig: OtMSQIT Example Quantum Circuit}
	\end{figure}
\end{tcolorbox}

\begin{tcolorbox}
	[
		grow to left by = 0.00 cm,
		grow to right by = 0.00 cm,
		colback = white,			
		enhanced jigsaw,			
		sharp corners,
		boxrule = 0.01 pt,
		toprule = 0.01 pt,
		bottomrule = 0.01 pt
	]
	\begin{figure}[H]
		\centering
		\includegraphics[ scale = 0.44, angle = 90, trim = {0cm 0cm 0cm 0cm}, clip ]{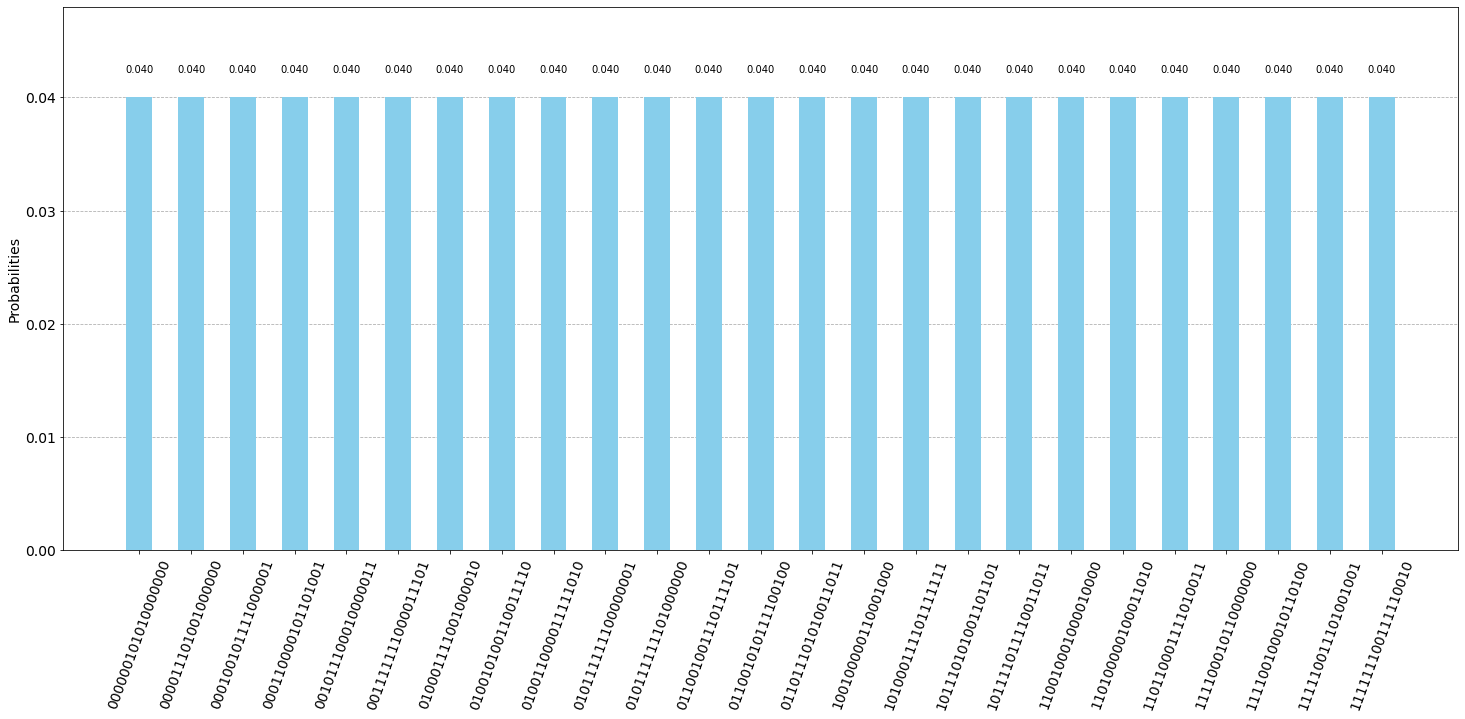}
		\caption{A few of the possible measurements and their corresponding probabilities for the circuit of Figure \ref{fig: OtMSQIT Example Quantum Circuit}.}
		\label{fig: OtMSQIT Example Measurement Outcomes}
	\end{figure}
\end{tcolorbox}

\section{Security analysis} \label{sec: Security Analysis}

The current section contains the security analysis of the OtMSQIT protocol. We proceed by assuming the existence of Eve, who is the cunning adversary that strives to compromise the security of the protocol and obtain some secret information like a PIV. As usual, we take for granted the existence of a classical authenticated channel, which will enable us to detect the presence of the eavesdropper Eve. We emphasize that the classical channel is not used for the transition of secret information; this privilege belongs exclusively to the quantum channel. The OtMSQIT protocol involves communication among $n$ parties, and relies on $\ket{ GHZ_{ n } }$ tuples, which makes it substantially more complex than typical QKD protocols involving only Alice, Bob, and Eve. Therefore, we provide an extensive and detailed security analysis in order to prove that it is information-theoretically secure. When considering strategies that may be employed by Eve, we often distinguish subcases depending on whether acts upon just one qubit or all $n - 1$ qubits from each $\ket{ GHZ_{ n } }$ tuple, so as to account for all possibilities. This accounts for the rather lengthy and technical current section. For a recent comprehensive text analyzing security issues of quantum protocols in general, we refer to \cite{Wolf2021} and the more recent \cite{Renner2023}.

At the end of the day, the security analysis of not just the OtMSQIT protocol, but of every quantum protocol relies on certain well-understood assumptions. We briefly state them for the purpose of making the current work self-contained. Naturally, we assume that quantum theory is correct, which in turn means that hallmark features such as the no-cloning theorem \cite{wootters1982single}, the monogamy of entanglement \cite{coffman2000distributed}, and nonlocality \cite{brunner2014bell} are valid. The unique features and enhanced efficiency of the quantum protocols are precisely due to these properties, otherwise, they would not offer any advantage over classical protocols. Secondly, we assume that quantum theory is complete, which implies that Eve is bound by the laws of quantum mechanics, and she cannot obtain more information beyond what these laws permit.


The importance of the validation test cannot be overestimated. If the test result is considered a failure, then the OtMSQIT protocol must be aborted. The secret embedding stage can safely begin only after the validation test has been successfully completed. The test itself consists of the following steps.

\begin{enumerate} [ left = 1.00 cm, labelsep = 1.50 cm ]
	\renewcommand\labelenumi{(\textbf{VT}$_\theenumi$)}
	\item	Alice communicates to every one of her agents Agent$_{ 0 }$, \dots, Agent$_{ n - 2 }$
	the positions of the decoys, so that they can measure them in the Hadamard basis.
	\item	Each agents sends back to Alice the results of her measurements. It is important to realize that the expected measurement outcome is, in general, different for every agent because, according to (\textbf{EDV}$_{ 2 }$), each qubit of the decoy tuple is prepared independently of the other qubits of the same tuple.
	\item	Alice analyzes the results received from her agents, and decides whether the test was successful or not, according to the following rationale.
	\begin{itemize}
		\item[$\diamond$]	If $0$ or very few wrong measurement outcomes are found, then Alice considers the validation test successful.
		\item[$\diamond$]	If the number of errors is
		$\approx \frac { d } { 4 }$, or above a similar threshold, then Alice deems that the validation test failed, in which case she aborts and terminates the protocol.
	\end{itemize}
\end{enumerate}

In the ideal scenario, where there is no eavesdropping and the quantum channel is perfect, there will be $0$ wrong measurement outcomes. In a more realistic scenario, even when there is no eavesdropping, we anticipate a few errors due to channel imperfections, but the number of errors is expected to be
$\ll \frac { d } { 4 }$. To understand the rationale behind the validation procedure, let us consider Eve's possible actions during the distribution phase. First, we make the critical remark that Eve has no way of knowing the position of the decoys. Therefore, Eve must treat all tuples in an identical manner.

\begin{enumerate} [ left = 1.00 cm, labelsep = 1.50 cm ]
	\renewcommand\labelenumi{(\textbf{EA}$_\theenumi$)}
	\item	\textbf{Measure \& Resend}. Eve intercepts one or more qubits from each $\ket{ GHZ_{ n } }$ $n$-tuple during their transmission from Alice to her agents. After measuring the intercepted qubit(s), Eve sends them back to their intended recipient. We make the following observations.
	\begin{itemize}
		\item[$\diamond$]	By the act of measurement, Eve destroys the entanglement. In view of the fact that in order to embed AIV into the global state of the distributed circuit entanglement is absolutely necessary, the protocol will fail. Hence, it is imperative that Alice discovers the loss of entanglement and aborts the execution of the protocol.
		\item[$\diamond$]	First, we examine the scenario where Eve always uses the computational basis for her measurements. In this scenario, the probability
		that Eve measures one decoy qubit and gets the wrong outcome is $\frac { 1 } { 2 }$,
		since all the decoys are measured in the wrong basis, and the probability to obtain the wrong outcome in such a case is $\frac { 1 } { 2 }$. Consequently, the probability that Eve obtains the correct outcome is $\frac { 1 } { 2 }$. This last probability implies that if Eve measures a second qubit from the same tuple, the probability to get two correct outcomes is way smaller. So, if Eve intercepts and measures two or more qubits from the same tuple, she stands to gain nothing in case they belong to a $\ket{ GHZ_{ n } }$ tuple, while she risks increasing the number of errors each time they belong to a decoy tuple. Therefore, Eve, being rational, will only measure one qubit from each tuple.
		\item[$\diamond$]	Now we consider the scenario where Eve randomly chooses the measurement basis between the computational or the Hadamard basis with equal probability. In this situation, the probability
		that Eve measures one decoy qubit and gets the wrong outcome is given by $\frac { 1 } { 4 }$,
		since the probability that a decoy is measured in the wrong basis is $\frac { 1 } { 2 }$, and, even then, the probability to get the wrong outcome is $\frac { 1 } { 2 }$. Consequently, the probability that Eve obtains the correct outcome is $\frac { 3 } { 4 }$. For the same reasons that we explained above, Eve will only measure one qubit from each tuple.
	\end{itemize}
	\item	\textbf{Intercept \& Send Fake $\ket{ GHZ_{ n } }$ $n$-tuples}. Eve intercepts a number of qubits from every $\ket{ GHZ_{ n } }$ $n$-tuple during their transmission from Alice to her agents. This number may range from just $1$ to $n - 1$. Eve can't clone the intercepted qubits due to the no-cloning theorem, but it is conceivable that she has prepared her own $\ket{ GHZ_{ n } }$ tuples. This opens up the possibility to keep the intercepted qubits and forward her own in their place. Again, we make the following remarks.
	\begin{itemize}
		\item[$\diamond$]	By doing so, Eve tampers with the entanglement. The protocol will fail because at least one PIV will not be encoded into the entanglement. Again, it is crucial that Alice discovers the loss of entanglement and aborts the execution of the protocol.
		\item[$\diamond$]	Eve, even if she were successful, will fail to gain any information. This is because her qubits are not entangled with Alice's qubits. The latter is the unique source of information who embeds the PIVs to those registers that are entangled with her own.
		\item[$\diamond$]	The flaw in this scenario is once again that Eve has no way of knowing the position of the decoys. If Eve intercepts just one qubit from every tuple, she will, inadvertently, replace $d$ decoy qubits with her $\ket{ GHZ_{ n } }$ qubits. When, during the validation test, these are measured in the Hadamard basis, the probability to obtain the wrong outcome is $\frac { 1 } { 2 }$. This will produce approximately $\approx \frac { d } { 2 }$ errors that will be easily noticed by Alice. If Eve intercepts $k$ qubits from each tuple, the probability to get at least one wrong measurement in a decoy tuple is $\frac { 2^{ k } - 1 } { 2^{ k } }$, which will result in approximately $\approx d \frac { 2^{ k } - 1 } { 2^{ k } }$ errors. In addition to the increased number of errors, Alice will easily notice that for $k$ decoy qubits in every decoy tuple the measurement results from her agents are identical, instead of uniformly distributed as they should be, as ordained by (\textbf{EDV}$_{ 2 }$). Practically, this strategy has almost zero chances of success, since Alice will, undoubtedly, infer the presence of Eve.
	\end{itemize}
	\item	\textbf{Entangle with Ancilla Qubits \& Measure Later}. Eve intercepts one qubit from every $\ket{ GHZ_{ n } }$ $n$-tuple during their transmission from Alice to her agents. Now, instead of measuring or replacing the intercepted qubits, Eve entangles them with her ancilla qubits, and then forwards them to their intended recipient. Eve plans to wait until the protocol completes, before measuring her qubits, hoping to gain useful information. In this case, we stress the next points.
	\begin{itemize}
		\item[$\diamond$]	The result of Eve's actions is that, instead of having $m$ $\ket{ GHZ_{ n } }$ tuples distributed among Alice and her $n - 1$ agents, we end up with $m$ $\ket{ GHZ_{ n + 1 } }$ tuples evenly distributed among Alice, her $n - 1$ agents, and Eve. Eve, even if she were successful, will fail to gain any information. This is because in order to decipher even a single PIV, she will require the contents of Alice and her agents' registers.
		\item[$\diamond$]	Of course, by doing so Eve changes the entanglement. The protocol will fail for the same reason as above, i.e., to decipher even a single PIV, Alice and her agents will require the contents of Eve's register. Again, it is imperative that Alice discovers the loss of entanglement and aborts the execution of the protocol.
		\item[$\diamond$]	Like in all previous case, the decoys will enable Alice to infer the presence of Eve. Recall that Eve has no way of knowing the position of the decoys. If Eve intercepts just one qubit from every tuple, she will entangle $d$ decoy qubits with her ancilla qubits. When, during the validation test, these are measured in the Hadamard basis, the probability to obtain the wrong outcome is $\frac { 1 } { 2 }$. This will produce approximately $\approx \frac { d } { 2 }$ errors that will be easily noticed by Alice. If Eve intercepts $k$ qubits from each tuple, the probability to get at least one wrong measurement in a decoy tuple is $\frac { 2^{ k } - 1 } { 2^{ k } }$, which will result in approximately $\approx d \frac { 2^{ k } - 1 } { 2^{ k } }$ errors. In addition to the increased number of errors, Alice will easily notice that for $k$ decoy qubits in every decoy tuple the measurement results from her agents are identical, instead of uniformly distributed as they should be, as ordained by (\textbf{EDV}$_{ 2 }$). This policy too has practically zero chances of success.
	\end{itemize}
\end{enumerate}

The above security analysis demonstrates that by setting the error threshold at $\approx \frac { d } { 4 }$ the OtMSQIT protocol is information-theoretically secure. Let us also emphasize the fact that even if Eve successfully eavesdrops during the entanglement distribution phase, she will get no information whatsoever because no information has been encoded yet. However it is still possible that she may disrupt the execution of the protocol. The validation test is designed to detect such an interference and abort the protocol. In closing, we remark that in the eventuality where the protocol is aborted, the security measures are not up to the task at hand. Hence, first measures must be taken to enhance security and then the process can start all over again.

\section{Discussion and conclusions} \label{sec: Discussion and Conclusions}

In this article, we introduce a new entanglement-based protocol for one-to-many simultaneous secure quantum information transmission, which we call OtMSQIT for short. The characteristic property of the new protocol is its extensibility, as it can be seamlessly generalized to an arbitrary number of entities. The proposed entanglement-based protocol is completely distributed and is provably information-theoretically secure. There many quantum protocols that achieve secure information communication between two parties, but most of them can't be generalized to situations involving parallel information transmission to two or more parties. This is achieved by the special way the transmitted information is embedded in the entangled state of the system, one of the distinguishing features compared to previous protocols. The advantage of this method is that it is seamlessly extensible and can be generalized to a setting involving an arbitrary number of players. This is not only useful, but necessary, whenever one information source must transmit simultaneously different secret messages to many recipients, without the need to apply the same two party protocol many times sequentially.  Due to its relative complexity, compared to similar cryptographic protocols, as it involves communication among $n$ parties, and relies on $\ket{ GHZ_{ n } }$ tuples, we provide an extensive and detailed security analysis so as to prove that it is information-theoretically secure. In terms of the capabilities of modern quantum apparatus, the implementation of the proposed protocol does not present any difficulty because it only requires CNOT and Hadamard gates. An additional advantage is that the local quantum circuits are identical for all information recipients.

\bibliographystyle{ieeetr}
\bibliography{OtMSSQIT}

\end{document}